\documentclass[pre,aps,twocolumn]{revtex4-2}
\usepackage{amsfonts}
\usepackage{times}
\usepackage{amsmath, amsthm}
\usepackage[colorlinks=true,citecolor=blue]{hyperref}
\usepackage[final]{graphicx}
\usepackage{amssymb, algorithm, algpseudocode}
\algnewcommand\algorithmicforeach{\textbf{for each}}
\algdef{S}[FOR]{ForEach}[1]{\algorithmicforeach\ #1\ \algorithmicdo}
\usepackage{xcolor,comment}
\usepackage{mathtools}

\begin{document}

\title{Data-driven Reduced-order Modeling of a Multistable Viscoelastic Flow}

\author{Shaoxuan Chen}
\email{Corresponding author: shaoxuanchen@umass.edu}
\affiliation{Department of Mathematics and
Statistics, University of Massachusetts
Amherst, Amherst, MA 01003-4515, USA}

\author{Su Yang}
\affiliation{College of Mathematics and Statistics, Chongqing University, Chongqing 401331, China}

%\author{Wei Zhu}
%\affiliation{School of Mathematics, Georgia Institute of Technology, Atlanta, GA 30332, USA}

\author{Simon J. Haward}
\affiliation{Okinawa Institute of Science and Technology Graduate University, Onna-son, Okinawa 904-0495, Japan}

\author{Amy Q. Shen}
\affiliation{Okinawa Institute of Science and Technology Graduate University, Onna-son, Okinawa 904-0495, Japan}

\author{Panayotis G.~Kevrekidis}
\affiliation{Department of Mathematics and
Statistics, University of Massachusetts
Amherst, Amherst, MA 01003-4515, USA}

\date{\today}

\begin{abstract}
    Viscoelastic flow past two side-by-side microcylinders has been shown to exhibit
geometry-dependent transitions among symmetric, converging, diverging, and asymmetric-diverging flow states, together with regions of
bistability and tristability. 
%Whether such complex bifurcation structures can be inferred from sparse measurements of stable flow states, without access to transient dynamics, remains an important challenge. 
Here, we develop a data-driven framework for
learning a two-dimensional, parameter-dependent normal form for
this bifurcation structure from sparse measurements of stable flow states. We use the flow asymmetry parameters \(I'\) and \(I''\) as state variables, with the gap ratio \(G\) and Weissenberg number \(Wi\) as control parameters. In the absence of trajectory data, the measured steady
states are enforced as stable equilibria of the learned
vector field through equilibrium and local-stability
constraints. We examine both polynomial and neural-network representations of the associated normal form and incorporate the
physical reflection symmetry directly
into the model.
%This inclusion
%ensures the correct relation between the two asymmetric-diverging
%states throughout the state and parameter domains. 
The resulting models recover the main geometry-dependent transitions and the overall structure of the experimental phase diagram. To reduce additional stable equilibria that are not constrained by the measurements, we introduce an iterative negative-sample procedure that refines the predicted equilibrium structure. {Combining equilibrium data, local
stability, physical symmetry, and negative samples yields the closest agreement with the experimental phase diagram among the models considered.} The remaining discrepancies reflect the limited information available, solely in the form of sparse equilibrium measurements. Overall, the framework aims to showcase a prototypical example of both the potential and the limitations of inferring nonlinear flow dynamics from sparse equilibrium measurements and highlight the importance of physical constraints and additional dynamical information for reconstructing complex bifurcation structures.

\end{abstract}

\maketitle

\section{Introduction}
\label{sec:introduction}

Viscoelastic flows \cite{groisman1997solitary,pan2013nonlinear,
sanchez2022understanding} at small length scales can exhibit instabilities
even when inertia is negligible~\cite{RevModPhys.77.977}. In this
regime, elastic stresses provide a source of nonlinearity that can
lead to symmetry breaking, time-dependent behavior, and the selection
of preferred flow paths. Flow-path selection is particularly relevant to
viscoelastic flows through porous media, where
the arrangement of obstacles can influence both
preferential flow pathways and the onset and
nature of elastic instabilities
~\cite{Walkama2020,Haward2021}. Understanding how geometry influences viscoelastic flow behavior is relevant to applications such as filtration \cite{miele2019stochastic,miele2026flow}, groundwater remediation \cite{budania2023comprehensive,obiri2014overview,wei2022recent}, and
enhanced oil recovery~\cite{yang2021micromodel,wei2014oil,liu2019pore}.

In recent years, data-driven reconstruction of governing equations has emerged as 
an avenue towards reduced-order modeling of nonlinear systems.
For instance, the sparse identification of nonlinear dynamics (SINDy) extracts
parsimonious ordinary differential equations from time-resolved
measurements by selecting ---through regression--- 
suitable terms from a prescribed library
\cite{Brunton2016SINDy}. Its extensions to partial differential
equations (PDEs) and weak-form formulations have enabled the identification
of spatiotemporal laws with improved robustness to measurement
noise~\cite{Rudy2017PDEFind,Messenger2021WeakSINDy}. Combining sparse
regression with neural representations has further enabled the
simultaneous discovery of effective coordinates and governing
equations~\cite{Champion2019Coordinates}, including the recovery of
PDEs from scarce and noisy
observations~\cite{Chen2021ScarcePDE}. In parallel, physics-informed
neural networks (PINNs) incorporate known constraints
into forward and inverse
problems~\cite{Raissi2019PINNs,Karniadakis2021PIML}, neural ordinary
differential equations provide flexible continuous-time
parameterizations of dynamical vector fields~\cite{Chen2018NeuralODE},
and Fourier neural operators learn efficient surrogate solution
mappings for families of parameterized PDEs~\cite{Li2021FNO}. Recent developments employing delay
embeddings and hidden-state inference have also enabled interpretable
dynamical models to be reconstructed from partial
observations~\cite{Bakarji2023Partial,Stepaniants2024Partial}. These
approaches have demonstrated considerable potential, although they
generally rely on time-resolved measurements, derivative information,
or some prior knowledge of the governing dynamics.

These advances have had significant impact in fluid mechanics,
where high-dimensional flow fields can exhibit relatively
low-dimensional coherent dynamics~\cite{Brunton2020FluidML}.
Constrained sparse Galerkin regression has yielded physically
consistent reduced-order models for cylinder wakes and cavity
flows~\cite{Loiseau2018Galerkin}, while physics-informed learning has
enabled the reconstruction of otherwise inaccessible velocity and
pressure fields from flow visualizations~\cite{Raissi2020HiddenFluid}.
In the case of experimental flows that are incompletely observed,
combining sparse
regression with physical constraints has led to significant progress in the recovery of
governing equations and latent pressure and forcing fields from noisy
measurements of weakly turbulent flows~\cite{Reinbold2021Robust}.
Related weak-form approaches have also recovered governing fluid-mechanical
relations from highly turbulent numerical
data~\cite{Gurevich2024SPIDER}.
%, while recent combinations of sparse
%identification and recurrent neural networks have learned
%interpretable latent dynamics from sparse sensor measurements,
%including applications to turbulent flows~\cite{Gao2026SINDySHRED}.
In the context of complex and viscoelastic fluids, rheology-informed
neural networks have been developed to identify constitutive
parameters~\cite{Dabiri2023FractionalRhINN}, and physics-informed
approaches have enabled the reconstruction of stress fields and
discrimination among candidate viscoelastic constitutive equations
from limited velocity data~\cite{Thakur2024ViscoelasticNet}.

Of particular relevance to the present study are earlier experimental investigations of shear-thinning viscoelastic flows past confined obstacles. For flow past a single cylinder, increasing elastic effects can induce a steady symmetry-breaking transition, whereby the fluid preferentially selects one of the two available paths around the obstacle~\cite{10.1039/c8sm02099j}. More recently, the work of~\cite{PhysRevLett.126.054501}, including some of the present authors, considered the richer configuration of two side-by-side microcylinders. Introducing a second obstacle creates an additional flow pathway between the cylinders, whose competition with the two outer paths gives rise to a substantially more complex bifurcation structure. In particular, the experiments revealed symmetric, converging, diverging, and asymmetric-diverging flow states, together with regimes of bistability and tristability, depending on the system geometry and imposed flow rate. These observations, obtained using a wormlike micellar solution, provide the experimental foundation for the present investigation.

Motivated by the aforementioned developments in data-driven modeling and the rich experimentally observed phenomenology of the above-mentioned viscoelastic flow, our aim herein is to bridge the gap between experimentally accessible, yet dynamically sparse, measurements and the reconstruction of effective low-dimensional dynamical equations. More specifically, we seek to infer a unified, coupled, parameter-dependent normal form capable of reproducing the experimentally observed bifurcation structure and multistability of the viscoelastic flow. Unlike approaches that rely on time-resolved trajectories or estimates of temporal derivatives, our framework uses measurements of stable steady states alone, expressed in terms of two suitably defined flow-asymmetry parameters. We investigate both polynomial and neural-network representations of the underlying vector field, incorporating equilibrium and local-stability constraints, the physical reflection symmetry of the system, and targeted negative samples to discourage unsupported stable branches. Beyond reconstructing the observed flow-state and bifurcation diagrams, our objective is to assess both the potential and the intrinsic limitations of inferring effective nonlinear dynamics from such restricted experimental information, particularly the extent to which the global bifurcation structure can be determined without access to transient dynamical measurements.

Our presentation of the above-mentioned findings is structured as follows.
In section II, we present the geometry of the configuration and discuss its
experimental layout and possible states, as well as the effective state variables
that we use to diagnose them. In section III, we present our mathematical
and computational methods and the resulting findings for the system's normal
form and its comparison with the experimentally identified one of~\cite{PhysRevLett.126.054501}.
Finally, section IV summarizes our principal findings and highlights a number of directions for 
future study.

\section{Experimental and Geometric Setup}
\begin{figure*}[!htbp]
    \centering

    \begin{minipage}[c]{0.7\textwidth}
        \centering
        \includegraphics[width=\linewidth]
        {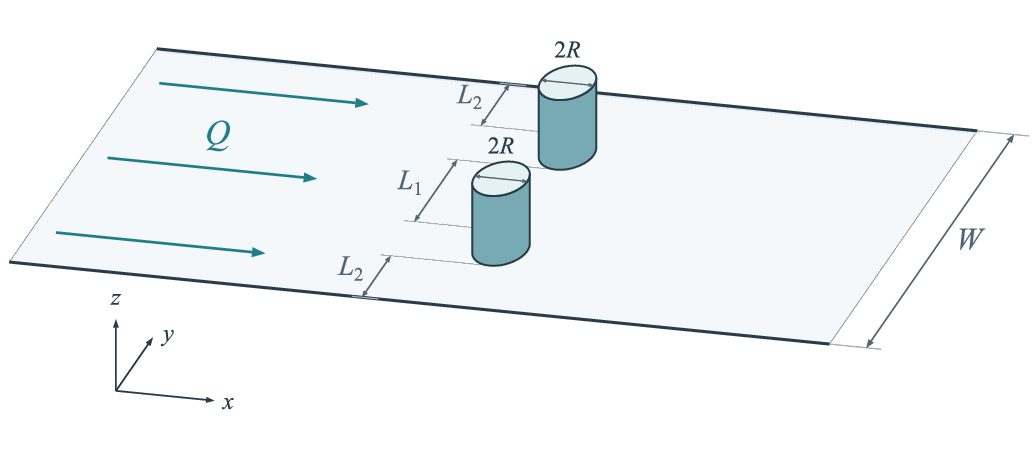}

        \vspace{0.3em}
        \textbf{(1)}
    \end{minipage}
    \hfill
    \begin{minipage}[c]{0.8\textwidth}
        \centering
        \includegraphics[width=\linewidth]
        {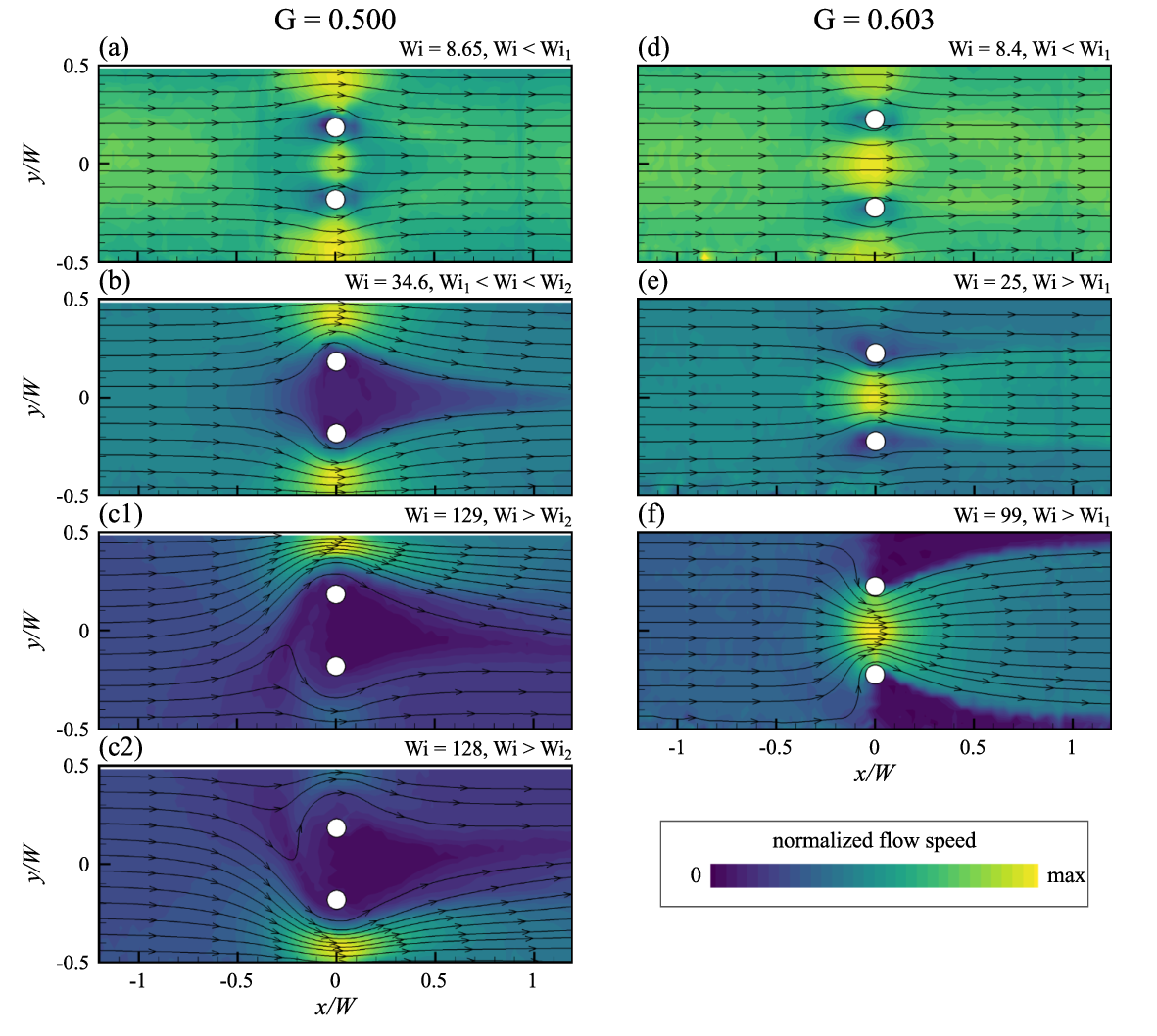}

        \vspace{0.3em}
        \textbf{(2)}
    \end{minipage}

    \caption{
Experimental geometry and representative flow states from
Ref.~\cite{PhysRevLett.126.054501}.
(1) Schematic of the channel in the \(x\)-\(y\) plane. The channel
width is \(W=400~\mu\mathrm{m}\), each cylinder has radius
\(R=20~\mu\mathrm{m}\), and the flow is directed from left to right
at volumetric rate \(Q\). The intercylinder and cylinder-wall gaps
have widths \(L_1\) and \(L_2\), respectively.
{(2) Evolution of the measured velocity fields with \(Wi\) for the wormlike micellar solution. Panels (a)--(c2) correspond to \(G=0.500\), where the first transition from the base-flow state to the diverging (D) state occurs at \(Wi_1\approx13\), followed by a second transition to the two reflection-related asymmetric-diverging
(AD) states at \(Wi_2\approx44\). Panels (d)--(f) correspond to \(G=0.603\), where the first transition from the base-flow state to the converging (C) state occurs at \(Wi_1\approx20\). No second steady-state transition is observed at this gap ratio, and the
C state persists at higher \(Wi\). Color indicates the normalized flow speed, and the black curves indicate streamlines. Reproduced from Hopkins et al.~\cite{PhysRevLett.126.054501}, with permission.}}
\label{fig:experimental_flow_overview}
\end{figure*}

The geometry of the experiment of~\cite{PhysRevLett.126.054501}
is shown in Fig.~\ref{fig:experimental_flow_overview}(1).
The two cylinders divide the channel into three gaps: an
intercylinder gap and two side gaps adjacent to the channel
walls. The relative gap widths are described by the geometric
parameter
\[
    G=\frac{L_1}{L_1+L_2},
\]
where \(L_1\) is the intercylinder separation and \(L_2\) is
the distance between each cylinder and the adjacent wall.
The imposed flow rate is expressed through the second key
parameter of our intended model, namely the Weissenberg number,
\[
    Wi=\frac{\lambda U}{R},
\]
where \(\lambda\) is the fluid relaxation time, \(R\) is
the cylinder radius, and \(U=Q/(WH)\) is the mean velocity
in a channel of width \(W\) and height \(H\), driven at
volumetric flow rate \(Q\).

{At low \(Wi\), prior to the onset of elastic instabilities,
the flow remains in its base-flow state, with fluid passing
through all three gaps. As \(Wi\) increases, the flow undergoes
a first transition at a critical Weissenberg number \(Wi_1\)
to either a diverging (D) or a converging (C) state, depending
on \(G\). As illustrated by the representative velocity fields
in part (2) of Fig.~\ref{fig:experimental_flow_overview}, for
\(G=0.500\), the flow first transitions to the D state,
in which the fluid avoids the intercylinder gap and passes
through the two side gaps. A second transition at \(Wi_2\)
leads to two asymmetric-diverging (AD) states, in which
the fluid preferentially selects either the upper or lower
side gap. In contrast, for \(G=0.603\), the first transition
leads to the C state, in which the fluid passes preferentially
through the intercylinder gap, and this state persists at
higher \(Wi\). These observations show that the flow can undergo different transitions depending on the geometric parameter \(G\), with multiple stable flow states possible under the same conditions.}

Accordingly, the flow states were characterized 
in~\cite{PhysRevLett.126.054501} by two dimensionless asymmetry parameters,

\begin{equation}
\begin{aligned}
    I'
    &=
    \frac{
        \frac{1}{2}\left(\bar{u}_{+}+\bar{u}_{-}\right)-\bar{u}_{0}
    }{
        \frac{1}{2}\left(\bar{u}_{+}+\bar{u}_{-}\right)+\bar{u}_{0}
    },\\[4pt]
    I''
    &=
    \frac{
        \bar{u}_{+}-\bar{u}_{-}
    }{
        \bar{u}_{+}+\bar{u}_{-}+\bar{u}_{0}
    },
\end{aligned}
\label{eq:viscoelastic_asymmetry}
\end{equation}
where \(\bar{u}_{+}\), \(\bar{u}_{-}\), and \(\bar{u}_{0}\) denote the average streamwise velocities through the upper, lower, and intercylinder gaps, respectively. Thus, \(I'\) measures the distribution of flow between the two side gaps and the center gap, whereas \(I''\) measures the asymmetry between the upper and lower side-gap flows. {
For visualization, we also consider the combined
asymmetry measure \(I=I'+I''\), which is shown in
the bifurcation diagrams but is not treated as an
independent state variable in the learned model.
}The symmetric state has \(I'\approx I''\approx0\), the D state has \(I'>0\) and \(I''\approx0\), the C state has \(I'<0\) and \(I''\approx0\), and the two AD states have \(I'>0\) and opposite nonzero values of \(I''\). Clearly, these diagnostics are well-positioned
to quantify and distinguish between the different possible states of
this dynamical system.

The steady-state bifurcation measurements reported by
Ref.~\cite{PhysRevLett.126.054501} show that the observed
flow transitions depend strongly on the gap ratio. For \(G=0.500\),
increasing \(Wi\) along the experimentally observed branches produces
the sequence symmetric--D--AD. In contrast, for \(G=0.603\), the symmetric state transitions to C, with no secondary steady symmetry-breaking transition observed within the steady-flow regime. At the intermediate value \(G=0.588\), both the D and C branches are
experimentally accessible over part of the parameter range. At larger
\(Wi\), the C branch coexists with the two reflection-related AD
branches, resulting in tristability.

The first transition from the symmetric state to either D or C is
quantified by \(I'\) [positive or negative, respectively], which measures the redistribution of flow
between the center and side gaps. The second transition from D to AD
is quantified by \(I''\) and breaks the upper--lower reflection
symmetry. The experimentally observed steady states constitute the
data used to constrain the equilibrium structure of our model. They
are shown as black open circles together with the model predictions in
Fig.~\ref{fig:polynomial_final_branches} and the subsequent
bifurcation diagrams. The flow-state diagram reported in
Ref.~\cite{PhysRevLett.126.054501}, including the D--C bistable and
AD--C tristable regions at intermediate \(G\), is reproduced in
panel~(B) of Fig.~\ref{fig:polynomial_final_phase_diagram} and the subsequent
phase-diagram comparisons. The individual transitions were described in
Ref.~\cite{PhysRevLett.126.054501} using fourth-order Landau-type
potentials fitted separately to \(I'\) and \(I''\). These fits capture
the local form of the observed bifurcations and describe how geometric
imperfections favor one branch over another, but they do not provide a
single (time-dependent) dynamical system for the coupled evolution of \(I'\) and \(I''\)
over the full \((G,Wi)\) parameter domain.

In the present work, we instead learn a low-dimensional, parameter-dependent normal form for the coupled dynamics of \(I'\) and \(I''\), with \(G\) and \(Wi\)
serving as control parameters. The learned vector field is then used to reconstruct the equilibrium branches and the phase diagram. This setting provides a test of how equilibrium data and physical constraints can be combined to infer a dynamically consistent global bifurcation structure from sparse experimental observations.

\section{Methods and Results}
\label{sec:methods_results}

We first introduce the learning problem and the equilibrium and stability
constraints shared by all model representations. We then describe the
reflection symmetry of the physical system and two approaches for
incorporating it into the learned dynamics. The polynomial and
neural-network models, together with their respective equilibrium
solvers, are presented in the subsequent subsections.

\subsection{General setting}
\label{sec:general_setting}

Following Eq.~\eqref{eq:viscoelastic_asymmetry}, we take
\begin{equation}
    x=I',
    \qquad
    y=I''
    \label{eq:model_notation}
\end{equation}
as the state variables, while \(G\) and \(Wi\) serve as the control
parameters. The experimental data obtained from
the work of~\cite{PhysRevLett.126.054501} are represented by
\begin{equation}
    \mathcal{D}
    =
    \left\{
        \left(
            x^i,y^i,G^i,Wi^i
        \right)
    \right\}_{i=1}^{N},
    \label{eq:equilibrium_dataset}
\end{equation}
where \((x^i,y^i)\) is an experimentally observed stable flow state
at \((G^i,Wi^i)\). In the bistable and tristable regimes, multiple
stable states may be associated with the same parameter values, and
each observed state is included as a separate sample in
\(\mathcal{D}\). Our objective is to learn a
parameter-dependent dynamical system
\begin{equation}
\left\{
\begin{aligned}
    \dot{x}
    &=
    f_1(x,y,G,Wi),\\
    \dot{y}
    &=
    f_2(x,y,G,Wi),
\end{aligned}
\right.
\label{eq:learned_vector_field}
\end{equation}
whose stable equilibria reproduce the measured branches over the
\((G,Wi)\) parameter domain.

At fixed \((G,Wi)\), an equilibrium of the learned system satisfies
\begin{equation}
    f_1(x,y,G,Wi)=0,
    \qquad
    f_2(x,y,G,Wi)=0.
    \label{eq:learned_equilibria}
\end{equation}
Because the measurements consist primarily of stable equilibria rather
than complete trajectories, we train the model by penalizing the vector
field residual at the observed states:
\begin{equation}
    \begin{aligned}
    &\mathcal{L}_{\mathrm{eq}}
    = \\
    &\frac{1}{N}\sum_{i=1}^{N}
    \left[
        \left|f_1(x^i,y^i,G^i,Wi^i)\right|^2
        +
        \left|f_2(x^i,y^i,G^i,Wi^i)\right|^2
    \right].
    \label{eq:equilibrium_loss}
    \end{aligned}
\end{equation}
Their local stability is determined by the Jacobian
\begin{equation}
    J_i
    =
    \begin{pmatrix}
        \partial_x f_1 & \partial_y f_1\\
        \partial_x f_2 & \partial_y f_2
    \end{pmatrix}_{(x^i,y^i,G^i,Wi^i)}.
    \label{eq:jacobian}
\end{equation}
At an equilibrium, the eigenvalues of \(J_i\) determine its local
linear stability. For a two-dimensional system, both eigenvalues have
negative real parts when
\begin{equation}
    \operatorname{tr}(J_i)<0,
    \qquad
    \det(J_i)>0.
    \label{eq:stability_conditions}
\end{equation}
These stability conditions are imposed through the loss
\begin{equation}
\begin{aligned}
\mathcal{L}_{\mathrm{stab}}
 = \\
\frac{1}{N}\sum_{i=1}^{N}\Big[
&\max\{0,\operatorname{tr}(J_i)+\alpha\}^{2}+w_{\det}\max\{0,\beta-\det(J_i)\}^{2}
\Big],
\end{aligned}
\label{eq:stability_loss}
\end{equation}
where \(\alpha,\beta>0\) provide stability margins,
and \(w_{\det}>0\) controls the relative weight of
the determinant penalty.

After training, we solve Eq.~\eqref{eq:learned_equilibria} throughout
the \((G,Wi)\) parameter domain. The polynomial equilibrium equations
are solved using Bertini~\cite{BHSW06}, whereas the neural-network
equilibrium equations are solved using a multi-start Newton-type root
solver. Details of the two numerical procedures are provided in the
corresponding subsections. At each parameter pair, the
Jacobian is evaluated at every equilibrium solution, and only the
stable equilibria satisfying Eq.~\eqref{eq:stability_conditions} are
used to construct the phase diagram. Based on the visual separation of the experimental branches, we use
the threshold \(\delta=0.15\) to assign each stable equilibrium to a
flow state. This threshold is used only for state classification and
does not enter the model training. 

In line with the discussion above,
and using the tolerance threshold $\delta$,  an equilibrium is classified as
symmetric when \(|x|<\delta\) and \(|y|<\delta\), as D when
\(x>\delta\) and \(|y|<\delta\), and as C when
\(x<-\delta\) and \(|y|<\delta\). The AD states are classified as
AD\(^{+}\) when \(x>\delta\) and \(y>\delta\), and as AD\(^{-}\)
when \(x>\delta\) and \(y<-\delta\). For the symmetry-constrained models, the threshold may not
distinguish weakly asymmetric states from the D state.
We therefore also check whether two distinct stable roots
initially classified as D have nearly equal \(x\) values
and opposite nonzero \(y\) values. If so, they are classified
as AD\(^{+}\) and AD\(^{-}\), respectively.
The numerical criteria for identifying such pairs are
given in Appendix~\ref{app:neural_network_implementation}. At a given \((G,Wi)\), coexistence of AD\(^{+}\) and AD\(^{-}\)
defines the AD-bistable regime. Coexistence of D and C defines the
D--C bistable regime, while coexistence of C with both AD states
defines the AD--C tristable regime. Applying this procedure throughout
the parameter domain produces the predicted stable equilibrium
branches and phase diagram.

\subsection{Reflection symmetry as a physical prior}
\label{sec:physical_information}

In the idealized geometry, the channel is symmetric with respect to
reflection across its horizontal centerline. This reflection exchanges
the upper and lower flow paths while leaving the intercylinder flow
unchanged. It therefore leaves \(I'\) unchanged and reverses the sign
of \(I''\). In terms of the model coordinates, the reflection is
\begin{equation}
    (x,y)\longmapsto(x,-y).
    \label{eq:state_reflection}
\end{equation}

Although small experimental imperfections may weakly break this symmetry, we use the reflection symmetry of the idealized geometry as a physical prior for the learned dynamics. The corresponding reflection-equivariance
conditions on the vector field are
\begin{equation}
\begin{aligned}
    f_1(x,-y,G,Wi)
    &=
    f_1(x,y,G,Wi),\\
    f_2(x,-y,G,Wi)
    &=
    -f_2(x,y,G,Wi).
\end{aligned}
\label{eq:reflection_parity_even_odd}
\end{equation}

Thus, \(f_1\) is even in \(y\), whereas \(f_2\) is odd in \(y\).
In particular, \(f_2(x,0,G,Wi)=0\), making \(y=0\) an invariant
subspace that contains the symmetric, D, and C branches. For a model
that satisfies Eq.~\eqref{eq:reflection_parity_even_odd} exactly, every
off-axis equilibrium \((x^\ast,y^\ast)\) has a reflected counterpart
\((x^\ast,-y^\ast)\) with the same local stability. These two
equilibria correspond to the AD\(^{+}\) and AD\(^{-}\) states.

We incorporate this physical information in two ways. The first is reflected-data augmentation, for which every
observation \((x^i,y^i,G^i,Wi^i)\) is supplemented by its reflected
counterpart \((x^i,-y^i,G^i,Wi^i)\):
\begin{equation}
    \mathcal{D}_{\mathrm{aug}}
    =
    \mathcal{D}
    \cup
    \left\{
        \left(
            x^i,-y^i,G^i,Wi^i
        \right)
    \right\}_{i=1}^{N}.
    \label{eq:augmented_dataset}
\end{equation}
The reflected state is assigned the same equilibrium and stability
constraints as the original observation. This data-level approach
encourages reflection-consistent predictions near the training data,
but it does not enforce exact symmetry throughout the domain.

The second approach incorporates
Eq.~\eqref{eq:reflection_parity_even_odd} directly into the model
representation. As a result, \(f_1\) is even and \(f_2\) is odd in
\(y\) throughout the state and parameter domains. The implementations
of these data-level and model-level approaches are described below for
the polynomial and neural-network models.

\subsection{Polynomial normal form}
\label{sec:polynomial_normal_form}

We first consider a polynomial representation of the
parameter-dependent vector field. 
This is motivated by the standard theory of normal forms
within dynamical systems~\cite{Strogatz2024,GuckenheimerHolmes1983}.
The polynomial model provides an
interpretable baseline for examining how the equilibrium, stability,
and symmetry constraints determine the learned bifurcation structure.

The available data consist of a sparse collection of experimentally
observed stable equilibria over the \((G,Wi)\) parameter domain. Our
objective is to learn a vector field whose stable equilibria reproduce
the observed branches and recover the bifurcation structure over the experimental parameter range. For the symmetry-constrained polynomial model presented below,
we use all available original equilibrium observations in
\(\mathcal{D}\), without reflected-data augmentation.
To improve numerical conditioning, all four inputs are scaled
according to
\begin{equation}
    \widetilde{z}
    =
    \frac{z}{s_z},
    \qquad
    s_z=\max_i |z^i|+10^{-12},
    \qquad
    z\in\{x,y,G,Wi\}.
    \label{eq:polynomial_scaling}
\end{equation}
The polynomial vector field and its stability constraints
are evaluated in these scaled coordinates.
For equilibrium reconstruction, the learned equations are
expressed in the original coordinates, as described below.

\begin{equation}
    \mathbf{X}
    =
    \begin{bmatrix}
        \widetilde{x}^{\,1} & \widetilde{y}^{\,1}
        & \widetilde{G}^{\,1} & \widetilde{Wi}^{\,1}\\
        \widetilde{x}^{\,2} & \widetilde{y}^{\,2}
        & \widetilde{G}^{\,2} & \widetilde{Wi}^{\,2}\\
        \vdots & \vdots & \vdots & \vdots\\
        \widetilde{x}^{\,N} & \widetilde{y}^{\,N}
        & \widetilde{G}^{\,N} & \widetilde{Wi}^{\,N}
    \end{bmatrix}
    \in\mathbb{R}^{N\times4}.
    \label{eq:polynomial_input_matrix}
\end{equation}

Naturally, such a representation bears some limitations.
Normal forms are expected to be valid in the vicinity of bifurcation
points, while here we seek to optimize the application of the identified
polynomial form throughout the datasets at hand. We expect this to partially
limit the suitability of the representation, yet, if successful, this may
enable us to obtain an interpretable representation not only of the immediate
vicinity of the bifurcation point but also of higher order effects. It is with that
in mind and being cognizant of these potential limitations that we propose
the relevant functional shape of the normal form terms.

From these inputs, we construct monomial features of the form
\begin{equation}
    \phi_k(\widetilde{x},\widetilde{y},
           \widetilde{G},\widetilde{Wi})
    =
    \widetilde{x}^{p_k}
    \widetilde{y}^{q_k}
    \widetilde{G}^{m_k}
    \widetilde{Wi}^{n_k}.
    \label{eq:polynomial_features}
\end{equation}

To control the complexity of the polynomial representation, we restrict
the degrees of the candidate monomials according to
\begin{equation}
   \begin{aligned}
    &p_k+q_k\leq d_{\mathrm{state}},
    \quad
    m_k+n_k\leq d_{\mathrm{param}},
    \quad \\
    &p_k+q_k+m_k+n_k\leq d_{\mathrm{total}},
    \label{eq:polynomial_degree_constraints}
    \end{aligned}
\end{equation}
where \(d_{\mathrm{state}}\), \(d_{\mathrm{param}}\), and
\(d_{\mathrm{total}}\) control the degree in the state variables, the
degree in the control parameters, and the total degree, respectively.

Separate dictionaries are constructed for the two components of the
vector field. To enforce the reflection symmetry described in Sec.~\ref{sec:physical_information}, the dictionary for \(f_1\) contains only monomials with even \(q_k\), while that for \(f_2\)
contains only monomials with odd \(q_k\). In the numerical experiments, we set
\[
    d_{\mathrm{state}}=4,
    \qquad
    d_{\mathrm{param}}=2,
    \qquad
    d_{\mathrm{total}}=5,
\]
and exclude the constant feature. The retained monomials for each component \(j\in\{1,2\}\)
are denoted by \(\{\phi_k^{(j)}\}_{k=1}^{M_j}\),
where \(M_j\) is the number of candidate features.
Evaluating these features at the scaled observations gives
\begin{widetext}
\begin{equation}
    \boldsymbol{\Phi}_j
    =
    \begin{bmatrix}
        \phi_1^{(j)}(\widetilde{x}^{\,1},\widetilde{y}^{\,1},
                     \widetilde{G}^{\,1},\widetilde{Wi}^{\,1})
        & \cdots &
        \phi_{M_j}^{(j)}(\widetilde{x}^{\,1},\widetilde{y}^{\,1},
                         \widetilde{G}^{\,1},\widetilde{Wi}^{\,1})\\
        \phi_1^{(j)}(\widetilde{x}^{\,2},\widetilde{y}^{\,2},
                     \widetilde{G}^{\,2},\widetilde{Wi}^{\,2})
        & \cdots &
        \phi_{M_j}^{(j)}(\widetilde{x}^{\,2},\widetilde{y}^{\,2},
                         \widetilde{G}^{\,2},\widetilde{Wi}^{\,2})\\
        \vdots & \ddots & \vdots\\
        \phi_1^{(j)}(\widetilde{x}^{\,N},\widetilde{y}^{\,N},
                     \widetilde{G}^{\,N},\widetilde{Wi}^{\,N})
        & \cdots &
        \phi_{M_j}^{(j)}(\widetilde{x}^{\,N},\widetilde{y}^{\,N},
                         \widetilde{G}^{\,N},\widetilde{Wi}^{\,N})
    \end{bmatrix}
    \in\mathbb{R}^{N\times M_j},
    \qquad j=1,2.
    \label{eq:polynomial_feature_matrices}
\end{equation}
\end{widetext}

Let
\[
    \mathbf{a}
    =
    [a_1,\ldots,a_{M_1}]^{\mathsf T},
    \qquad
    \mathbf{b}
    =
    [b_1,\ldots,b_{M_2}]^{\mathsf T}
\]
denote the learnable coefficient vectors. The vector-field values at the observed states are assembled through the linear projections
\begin{equation}
    \mathbf{f}_1
    =
    \boldsymbol{\Phi}_1\mathbf{a},
    \qquad
    \mathbf{f}_2
    =
    \boldsymbol{\Phi}_2\mathbf{b}.
    \label{eq:polynomial_matrix_model}
\end{equation}

Equivalently, the learned normal form in the scaled
coordinates is
\begin{equation}
\left\{
\begin{aligned}
    \dot{\widetilde{x}}
    &=
    f_1(\widetilde{x},\widetilde{y},
        \widetilde{G},\widetilde{Wi})
    =
    \sum_{k=1}^{M_1}
    a_k\phi_k^{(1)}
    (\widetilde{x},\widetilde{y},
     \widetilde{G},\widetilde{Wi}),\\
    \dot{\widetilde{y}}
    &=
    f_2(\widetilde{x},\widetilde{y},
        \widetilde{G},\widetilde{Wi})
    =
    \sum_{k=1}^{M_2}
    b_k\phi_k^{(2)}
    (\widetilde{x},\widetilde{y},
     \widetilde{G},\widetilde{Wi}).
\end{aligned}
\right.
\label{eq:polynomial_normal_form}
\end{equation}

Thus, the model is nonlinear in the state variables and control parameters but linear in the unknown coefficients. The coefficients are learned by imposing the equilibrium conditions at the observed states and enforcing their local stability through the Jacobian. Since the model is polynomial, the Jacobian is evaluated
analytically by differentiating the monomial features with
respect to \(\widetilde{x}\) and \(\widetilde{y}\), avoiding
numerical differentiation. To have a compact representation, we use the sparsity penalty
\begin{equation}
    \mathcal{L}_{\mathrm{sparse}}
    =
    \|\mathbf{a}\|_1+\|\mathbf{b}\|_1.
    \label{eq:polynomial_sparsity_loss}
\end{equation}
The complete training objective is
\begin{equation}
    \mathcal{L}_{\mathrm{poly}}
    =
    \mathcal{L}_{\mathrm{eq}}
    +
    \lambda_{\mathrm{stab}}
    \mathcal{L}_{\mathrm{stab}}
    +
    \lambda_{\mathrm{sparse}}
    \mathcal{L}_{\mathrm{sparse}},
    \label{eq:polynomial_total_loss}
\end{equation}
where \(\mathcal{L}_{\mathrm{eq}}\) and \(\mathcal{L}_{\mathrm{stab}}\) are defined in Eqs.~\eqref{eq:equilibrium_loss} and~\eqref{eq:stability_loss}. The coefficients are optimized using full-batch Adam.
The training hyperparameters and additional implementation details are provided in
Appendix~\ref{app:polynomial_implementation}.

During model development, we also examined the role of
reflection symmetry in the polynomial model. The unrestricted model did not reproduce the experimental equilibrium branches
accurately, as shown in Appendix~\ref{app:polynomial_ablation}.
While reflected-data augmentation, as in Eq.~\eqref{eq:augmented_dataset}, improved the reconstruction, imposing the even--odd structure directly on the polynomial representation provided a closer match to the experimental branches. Furthermore, combining the symmetry constraint with data augmentation did not lead to any noticeable improvement. We therefore focus on the symmetry-constrained polynomial model without data augmentation in the results presented below.

After fitting the polynomial vector field to all available
equilibrium observations, we substitute the input scaling
to express the equilibrium equations in the original
\(x\), \(y\), \(G\), and \(Wi\) coordinates.
For a monomial with exponents \((p,q,m,n)\), the
corresponding coefficient becomes
\begin{equation}
    c_{\mathrm{eq},k}
    =
    \frac{c_{\mathrm{scaled},k}}
    {s_x^p s_y^q s_G^m s_{Wi}^n},
    \qquad k=1,2.
    \label{eq:polynomial_coefficient_conversion}
\end{equation}
This transformation preserves the equilibrium solutions. The equilibria are then obtained by solving
Eq.~\eqref{eq:learned_equilibria} in the original
input coordinates using Bertini version 1.7,
a numerical homotopy-continuation package for
polynomial equations~\cite{BHSW06}. Each returned finite real root is mapped to the scaled
coordinates and checked against the full learned
polynomial equations. A root is retained when
\begin{equation}
    \sqrt{
        |f_1(\widetilde{x},\widetilde{y},
             \widetilde{G},\widetilde{Wi})|^2+
        |f_2(\widetilde{x},\widetilde{y},
             \widetilde{G},\widetilde{Wi})|^2
    }
    <10^{-6}.
    \label{eq:polynomial_root_residual}
\end{equation}
Stability is determined from the Jacobian with respect to
\(\widetilde{x}\) and \(\widetilde{y}\), consistent with
the training procedure. Only stable roots satisfying
\(|x|\leq1.5\) and \(|y|\leq1.5\) in the original coordinates
are retained for the bifurcation and phase diagrams.

For the one-parameter bifurcation diagrams, the calculation
is repeated at \(G=0.500\), \(0.588\), and \(0.603\) for
\[
    Wi=1,6,\ldots,196.
\]
For the two-parameter phase diagram, the same calculation
is performed over
\[
    G=0.50,0.51,\ldots,0.63,
    \qquad
    Wi=8,13,\ldots,198.
\]
At each parameter pair, the retained stable equilibria are classified as symmetric, C, D, or AD states using the criteria introduced in Sec.~\ref{sec:general_setting}. The collection of coexisting stable states determines the reconstructed phase at that parameter value.
Thus, the bifurcation and phase diagrams are obtained from the
equilibrium structure of the learned vector field rather than from
direct prediction of phase labels.

\begin{figure*}[t!]
    \centering
    \includegraphics[width=.98\textwidth]
    {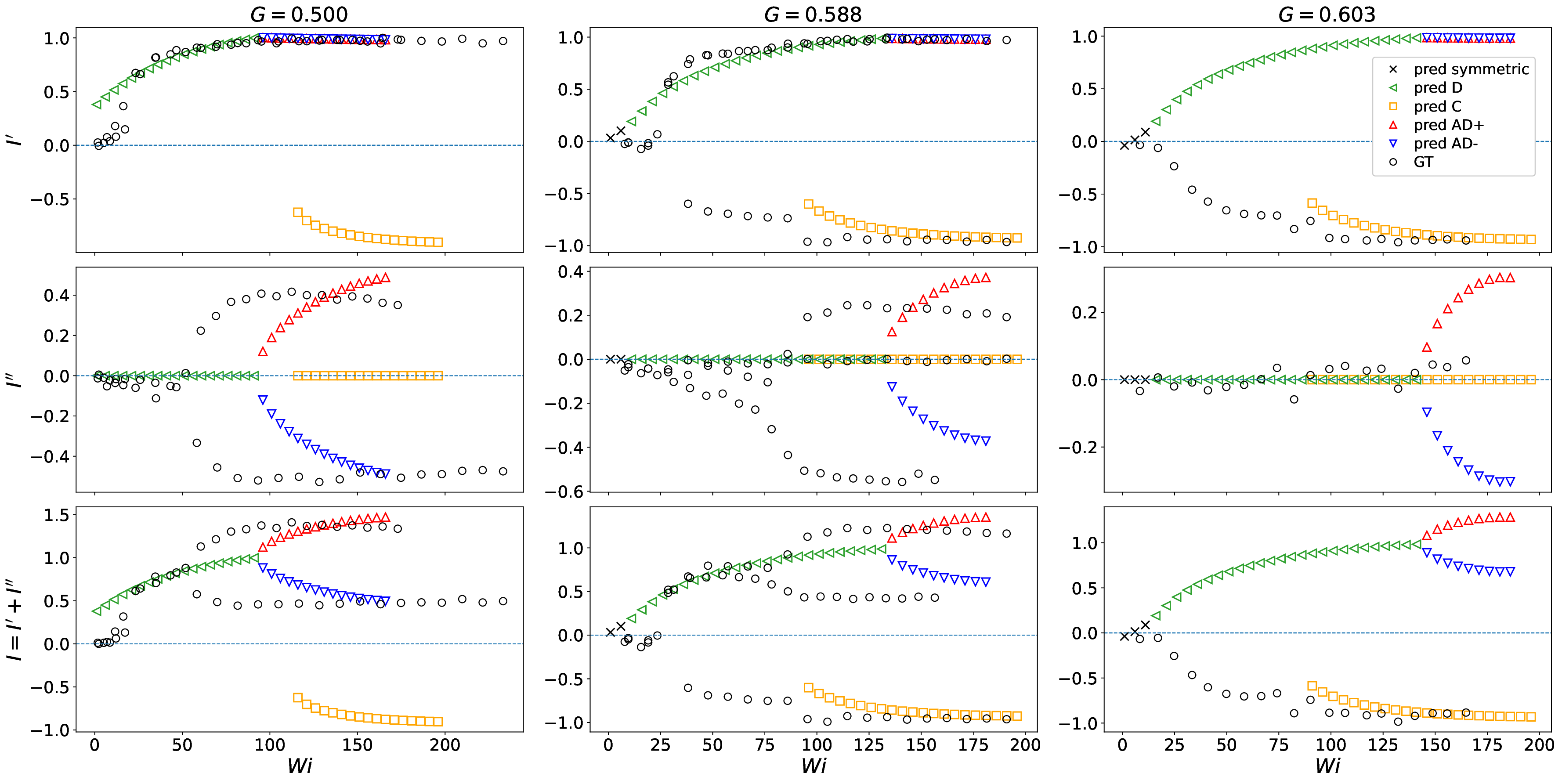}
    \caption{Bifurcation diagrams reconstructed from the final
symmetry-restricted polynomial normal form without reflected-data augmentation. The columns correspond,
from left to right, to \(G=0.500\), \(G=0.588\), and \(G=0.603\), while
the rows show \(I'=x\), \(I''=y\), and \(I=I'+I''\), respectively. Black open circles denote the experimental equilibrium measurements, labeled as "GT" (ground truth) in the legend.
Crosses denote the predicted symmetric states, left-pointing triangles the D states,
    squares the C states, and upward- and downward-pointing triangles
    the AD\(^{+}\) and AD\(^{-}\) states, respectively. Only stable
    equilibria of the learned vector field are shown. Predicted
    equilibria satisfying \(\lvert I'\rvert<0.15\) and
    \(\lvert I''\rvert<0.15\) are classified as symmetric. The model
    reproduces the main observed branches, although several
    transitions are shifted toward larger \(Wi\). It also predicts an
    additional C branch at \(G=0.500\) and additional D and AD branches
    at \(G=0.603\).
    }
    \label{fig:polynomial_final_branches}
\end{figure*}

\begin{figure*}[t!]
    \centering

    \begin{minipage}[t]{0.49\textwidth}
        \vspace{0pt}
        \centering
        \includegraphics[
            width=\linewidth,
            height=0.38\textheight,
            keepaspectratio
        ]{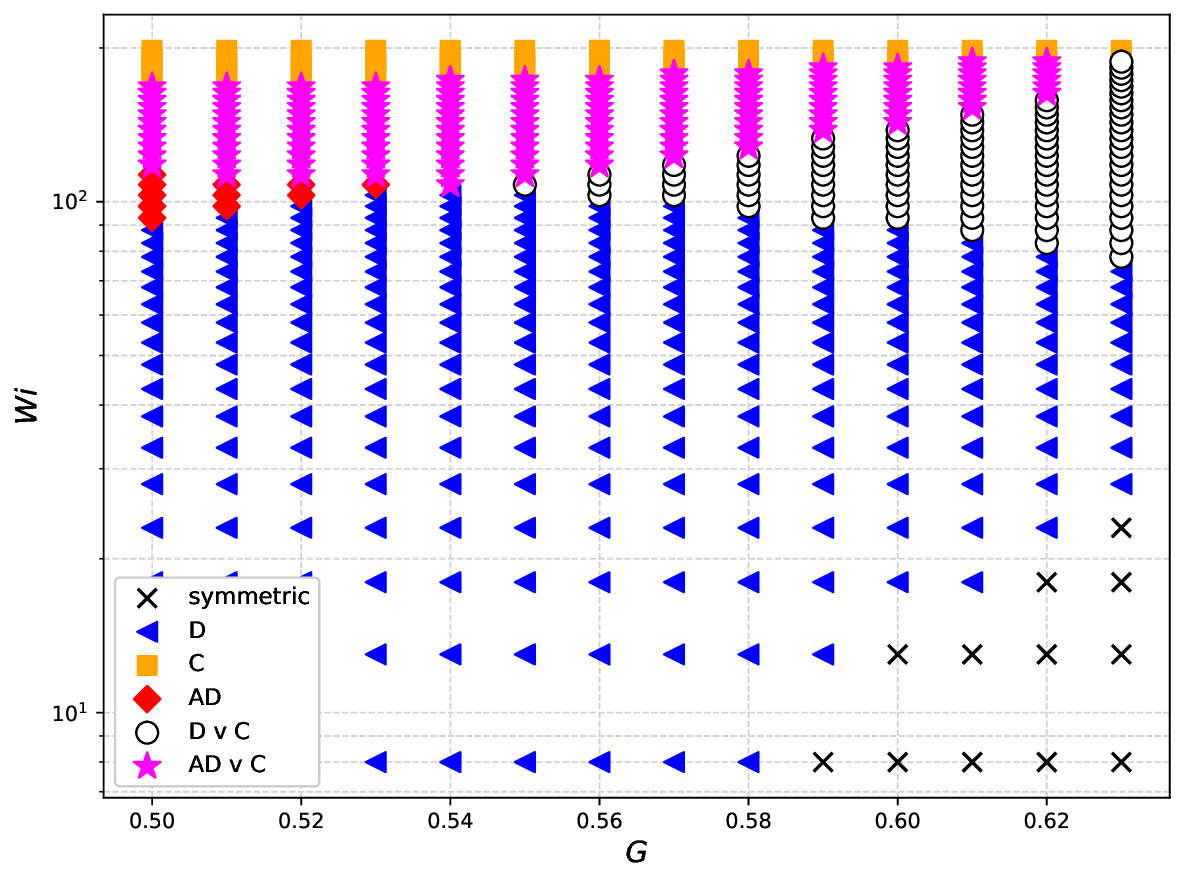}

        \vspace{0.3em}
        \textbf{(A)}
    \end{minipage}
    \hfill
    \begin{minipage}[t]{0.49\textwidth}
        \vspace{0pt}
        \centering
        \includegraphics[
            width=\linewidth,
            height=0.38\textheight,
            keepaspectratio
        ]{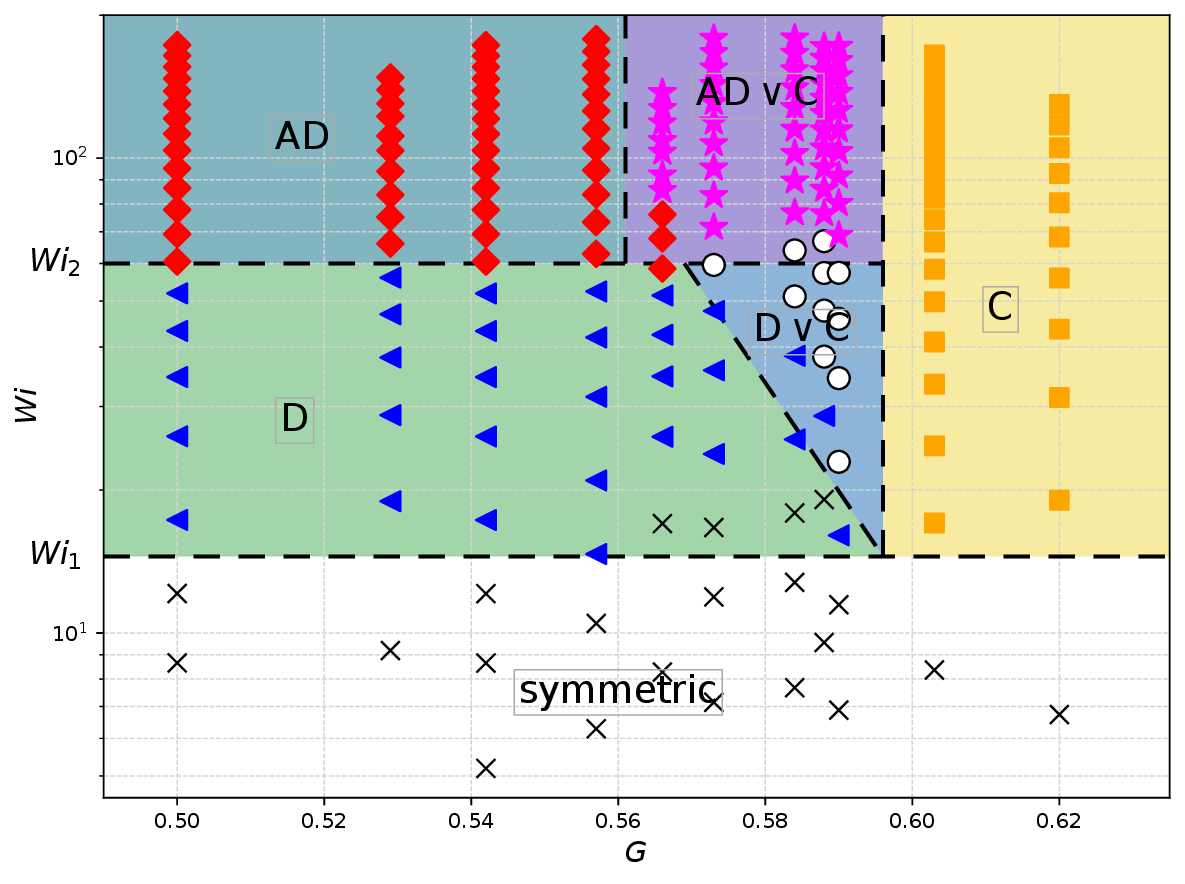}

        \vspace{0.3em}
        \textbf{(B)}
    \end{minipage}

    \caption{
    Comparison of the reconstructed and experimental phase diagrams.
    (A) Phase diagram obtained from the stable equilibria of the symmetry-constrained polynomial model without reflected-data augmentation. Crosses, left-pointing
    triangles, squares, and diamonds denote symmetric, D, C, and AD states, respectively. Open circles denote D--C bistability, and stars denote AD--C tristability. The model predicts D at low \(Wi\) over much of the parameter range and C at large \(Wi\) for
    almost all \(G\), resulting in overly broad multistable regions.
    (B) Experimental flow-stability diagram showing symmetric flow at
    low \(Wi\), D/AD behavior at small \(G\), C behavior at large
    \(G\), and multistability at intermediate \(G\). The latter
    panel is {reproduced from Hopkins et al.~\cite{PhysRevLett.126.054501}, with permission.}
    }
    \label{fig:polynomial_final_phase_diagram}
\end{figure*}

Figures~\ref{fig:polynomial_final_branches}
and~\ref{fig:polynomial_final_phase_diagram} present the bifurcation and phase
diagrams reconstructed from the final polynomial model with the exact
even--odd structure and no reflected-data augmentation. The model is fitted
using all available equilibrium observations and then evaluated over
the denser parameter grids described above. Figure~\ref{fig:polynomial_final_branches} compares the predicted
stable equilibria with the experimental observations at \(G=0.500\),
\(0.588\), and \(0.603\). The black open circles denote the
experimental data, while the crosses and colored symbols denote the
stable equilibria of the learned polynomial vector field.

For \(G=0.500\), the model reproduces the observed D branch and the two
AD branches. However, the transition from D to AD occurs at a larger
\(Wi\) than in the experiment. The model also predicts an additional
C branch for \(Wi\gtrsim100\), which is not observed experimentally.
For \(G=0.588\), the model recovers the D, C, and two AD branches, but
the predicted onsets of both the C and AD states are shifted toward
larger \(Wi\). In particular, the AD branches are absent over a
substantial part of the parameter range in which they are observed
experimentally. For \(G=0.603\), the model reproduces the C branch at large \(Wi\),
although its onset is again delayed. The model also predicts an
additional D branch over a broad range of \(Wi\). At larger \(Wi\),
this branch undergoes a further symmetry-breaking transition into two
AD branches. Neither the D nor the AD states are observed
experimentally at this gap ratio. Thus, while the polynomial model
captures the main types of stable branches, it is less accurate in
locating their onset and introduces additional stable states in parts
of the parameter domain.

Figure~\ref{fig:polynomial_final_phase_diagram} compares the
reconstructed phase diagram with the experimental result. In the
experiment, the low-\(Wi\) flow is symmetric for all \(G\). Increasing
\(Wi\) produces D and then AD at small \(G\), C at large \(G\), and
D--C and AD--C coexistence over a narrow intermediate range. The
polynomial model does not fully recover this dependence on \(G\).
Instead, D extends to the lowest sampled \(Wi\) for
\(G\lesssim0.58\), while C is predicted at sufficiently large \(Wi\)
over almost the entire range of \(G\). The additional C and D branches
produce broader than expected AD--C and D--C coexistence regions, respectively.

At small \(G\), the additional C branch produces coexistence with the
two AD branches over a broad high-\(Wi\) range. The model therefore
predicts tristability between C, AD\(^{+}\), and AD\(^{-}\) in a
region where the experiment shows only the two AD states. At larger
\(G\), the model predicts D--C bistability at intermediate \(Wi\),
followed by tristability between C, AD\(^{+}\), and AD\(^{-}\) at
higher \(Wi\). In contrast, the experiment identifies only the C state
in this part of the parameter domain. The predicted multistable
regions therefore extend well beyond the narrow intermediate-\(G\)
range observed experimentally. Although the symmetry-constrained polynomial model recovers the individual flow states, it does not reproduce their distribution over the full parameter domain. This may partly reflect the limited flexibility of the polynomial representation over a wide parameter range. This limitation motivates the more flexible neural-network representation
considered next.

\subsection{Neural-network normal forms}
\label{sec:neural_network_normal_forms}

The polynomial representation is interpretable, but its flexibility is
limited by the prescribed monomial library. We therefore consider
neural-network representations that allow more general nonlinear
dependence on the state variables and control parameters. We develop these models in three stages. We first
parameterize the vector field directly using an unconstrained neural
network. We then impose the physical reflection symmetry exactly through
a symmetry-projected architecture. Finally, we introduce an iterative
negative-sample procedure to remove additional stable equilibria that
are not supported by the experimental observations.

\subsubsection{Direct neural-network model}
\label{sec:direct_neural_network}
We first represent the vector field directly using a neural network. Unlike the polynomial model, where all four inputs are scaled to improve the numerical conditioning of the polynomial features, we scale only the control parameters \(G\) and \(Wi\) following Eq.~\eqref{eq:polynomial_scaling}. Since \(x\) and \(y\) are normalized flow-asymmetry measures with values typically between \(-1\) and \(1\), they can be used directly as network inputs without introducing large differences in input scales. The observed equilibria are then assembled into the input matrix

\begin{equation}
    \mathbf{X}_{\mathrm{NN}}
    =
    \begin{bmatrix}
        x^1 & y^1 & \widetilde{G}^{\,1} & \widetilde{Wi}^{\,1}\\
        x^2 & y^2 & \widetilde{G}^{\,2} & \widetilde{Wi}^{\,2}\\
        \vdots & \vdots & \vdots & \vdots\\
        x^N & y^N & \widetilde{G}^{\,N} & \widetilde{Wi}^{\,N}
    \end{bmatrix}
    \in\mathbb{R}^{N\times4}.
    \label{eq:direct_nn_input_matrix}
\end{equation}

The network consists of a shared nonlinear trunk
\(T_{\boldsymbol{\theta}}\) followed by two scalar output heads
\(H_{\boldsymbol{\theta},1}\) and
\(H_{\boldsymbol{\theta},2}\). Applied rowwise to the observations, the
architecture can be summarized as
\begin{equation}
    \begin{aligned}
    &\mathbf{X}_{\mathrm{NN}}
    \in\mathbb{R}^{N\times4}
    \xmapsto{\;T_{\boldsymbol{\theta}}\;}
    \mathbf{H}
    \in\mathbb{R}^{N\times50} \\
    &\xmapsto{\;
        H_{\boldsymbol{\theta},1},
        H_{\boldsymbol{\theta},2}
    \;}
    \left(
        \mathbf{f}_1,\mathbf{f}_2
    \right)
    \in\mathbb{R}^{N}\times\mathbb{R}^{N}.
    \label{eq:direct_nn_architecture}
    \end{aligned}
\end{equation}
The shared trunk contains three fully connected layers of width \(50\),
each followed by a hyperbolic tangent activation. Each output head
contains a fully connected layer of width \(64\), a hyperbolic tangent
activation, and a scalar output layer. The resulting vector field is
\begin{equation}
\left\{
\begin{aligned}
    \dot{x}
    &=
    f_{\boldsymbol{\theta},1}
    (x,y,\widetilde{G},\widetilde{Wi})
    =
    H_{\boldsymbol{\theta},1}
    \left[
        T_{\boldsymbol{\theta}}
        (x,y,\widetilde{G},\widetilde{Wi})
    \right],\\
    \dot{y}
    &=
    f_{\boldsymbol{\theta},2}
    (x,y,\widetilde{G},\widetilde{Wi})
    =
    H_{\boldsymbol{\theta},2}
    \left[
        T_{\boldsymbol{\theta}}
        (x,y,\widetilde{G},\widetilde{Wi})
    \right].
\end{aligned}
\right.
\label{eq:direct_neural_network}
\end{equation}
The shared trunk learns a common nonlinear representation of the flow
state, while the two output heads independently parameterize the two
components of the dynamics.

As in the polynomial formulation, the network is trained using stable equilibrium observations rather than time trajectories. The equilibrium loss follows Eq.~\eqref{eq:equilibrium_loss}, with the vector field evaluated at the scaled control parameters. The Jacobian derivatives are computed by automatic differentiation, and
the resulting trace and determinant are used in the stability loss
defined in Eq.~\eqref{eq:stability_loss}. Because the neural-network
representation does not have a polynomial coefficient vector, no
sparsity penalty is included. The complete training objective is
therefore
\begin{equation}
    \mathcal{L}_{\mathrm{direct}}
    =
    \mathcal{L}_{\mathrm{eq}}
    +
    \lambda_{\mathrm{stab}}
    \mathcal{L}_{\mathrm{stab}}.
    \label{eq:direct_nn_total_loss}
\end{equation}

After training, the equilibrium branches are reconstructed
by solving Eq.~\eqref{eq:learned_equilibria} using the scaled control parameters as neural-network inputs. Since the hyperbolic tangent network
is nonpolynomial, Bertini cannot be used for this calculation. Instead,
we use a multi-start Newton-type root-finding procedure based on the Powell
hybrid method, initialized from a grid of points in the \((x,y)\) domain. Converged solutions are filtered
according to their residuals, duplicate roots are merged, and local
stability is determined from the eigenvalues of the Jacobian. The same
root-finding procedure is used for the neural-network models introduced
below. Additional numerical details are provided in
Appendix~\ref{app:neural_network_implementation}.

\begin{figure*}[t!]
    \centering
    \includegraphics[width=\textwidth]
    {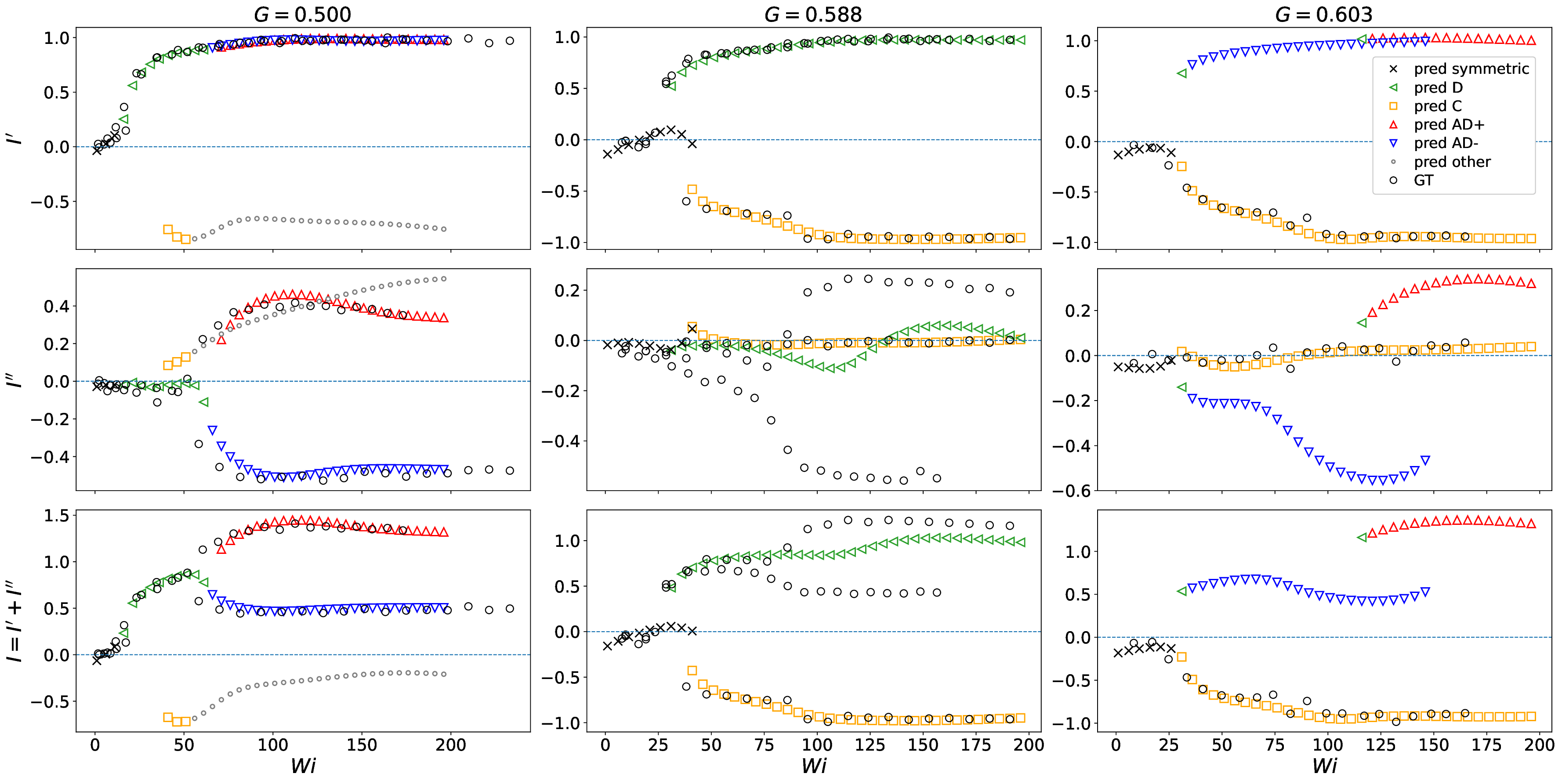}
    \caption{Bifurcation diagrams reconstructed from the direct
    neural-network model trained without reflected-data augmentation.
    The columns correspond to \(G=0.500\), \(G=0.588\), and
    \(G=0.603\), while the rows show \(I'\), \(I''\), and
    \(I=I'+I''\), respectively. Colored symbols denote stable
    equilibria predicted by the learned vector field, and black open
    circles denote the experimental measurements. The model delays the
    AD transition at \(G=0.500\), does not recover the AD branches at
    \(G=0.588\), and predicts additional stable equilibria at
    \(G=0.500\) and \(G=0.603\). The predicted AD\(^{+}\) and
    AD\(^{-}\) branches also do not satisfy the expected reflection
    symmetry.}
    \label{fig:direct_nn_no_augmentation_bifurcation}
\end{figure*}

The direct neural-network model is trained using the original equilibrium
observations without reflected-data augmentation.
Figure~\ref{fig:direct_nn_no_augmentation_bifurcation} compares the
stable equilibria predicted by the learned vector field with the
experimental measurements at three representative gap ratios. The
model reproduces several individual branches but does not recover the complete bifurcation structure. At \(G=0.500\), it captures the initial D branch
and the subsequent AD branches, although the predicted transition in
\(I''\) occurs at a larger \(Wi\) than in the experimental data. The
two predicted AD branches are not exact reflections of one another.
The model also predicts a short C branch at intermediate \(Wi\) and
additional unclassified equilibria at larger \(Wi\). At \(G=0.588\),
the model approximates the coexisting D and C branches but does not
recover the secondary transition from D to the two AD states. At
\(G=0.603\), the experimentally observed C branch is reproduced
reasonably well, but the model also predicts additional positive-\(I'\)
equilibria, most of which are classified as AD states. The predicted
AD\(^{+}\) and AD\(^{-}\) branches occur over different ranges of
\(Wi\) and do not satisfy the expected reflection symmetry.

\begin{figure*}[t!]
    \centering

    \begin{minipage}[t]{0.49\textwidth}
        \vspace{0pt}
        \centering
        \includegraphics[
            width=\linewidth,
            height=0.38\textheight,
            keepaspectratio
        ]{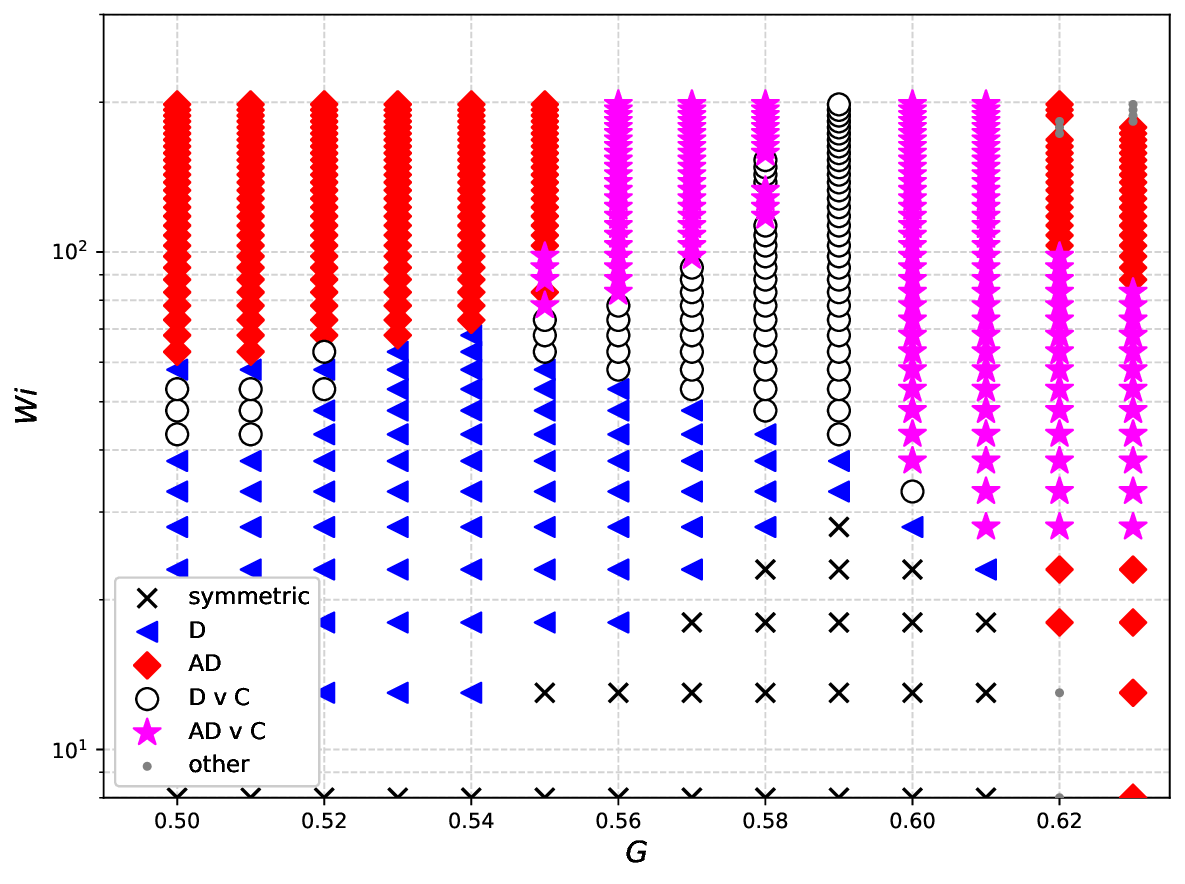}

        \vspace{0.3em}
        \textbf{(A)}
    \end{minipage}
    \hfill
    \begin{minipage}[t]{0.49\textwidth}
        \vspace{0pt}
        \centering
        \includegraphics[
            width=\linewidth,
            height=0.38\textheight,
            keepaspectratio
        ]{figures/flow_stability_diagram}

        \vspace{0.3em}
        \textbf{(B)}
    \end{minipage}

    \caption{Comparison of the reconstructed and experimental phase
    diagrams. (A) Phase diagram reconstructed from the stable
    equilibria of the direct neural-network model trained without
    reflected-data augmentation. The model broadly recovers the D and
    AD regions at small \(G\), but the predicted D--C and AD--C regions
    extend beyond the intermediate range of \(G\) observed
    experimentally. The model also fails to recover the C-only region
    at larger \(G\), where additional AD, multistable, and unclassified
    states remain. (B) Experimental flow-stability diagram in the
    \((G,Wi)\) parameter space. The symbols indicate the measured
    symmetric, D, C, and AD states, while the colored regions and dashed
    boundaries identify the corresponding stable and multistable
    regimes. The intermediate-gap region contains D--C bistability and
    AD--C tristability. {Reproduced from Hopkins et al.~\cite{PhysRevLett.126.054501}, with permission.}}
    \label{fig:direct_nn_no_augmentation_phase_diagram}
\end{figure*}

These branch-level errors are also reflected in the reconstructed phase
diagram shown in
Fig.~\ref{fig:direct_nn_no_augmentation_phase_diagram}(A). At small
\(G\), the model broadly captures the progression from the symmetric
state to D and then AD as \(Wi\) increases. However, it also predicts
D--C coexistence at gap ratios as small as \(G=0.50\), where the
experiments identify only D or AD states. The predicted AD--C region
similarly extends to approximately \(G=0.55\), beyond the
intermediate-\(G\) range observed experimentally. The differences become more pronounced at larger \(G\). Near
\(G=0.59\)--\(0.61\), the model predicts broad D--C and AD--C regions,
whereas the experimental diagram shows a transition from the narrow
multistable region to the C state. The model does not recover a C-only
region over the sampled parameter domain. Instead, additional AD,
AD--C, and unclassified states persist for \(G\gtrsim0.60\), where the
experiments predominantly identify C. Thus, the direct neural network
captures part of the experimental phase diagram at small \(G\) but does not
correctly reproduce the extent of the multistable regions or the
C-dominated regime at larger \(G\).

We also tested reflected-data augmentation during model development. It
did not improve the direct neural-network reconstruction and resulted in
a less accurate reconstruction of the stable equilibrium branches. We
therefore do not use reflected-data augmentation in the neural-network
models presented below. The results above also show that the equilibrium
and stability losses do not by themselves enforce reflection symmetry
in the learned vector field. We next impose the symmetry directly
through the network architecture, ensuring that \(f_1\) is even and
\(f_2\) is odd in \(y\) throughout the state and parameter domains.

\subsubsection{Symmetry-projected neural-network normal form}
\label{sec:symmetry_neural_network}

The direct neural network does not guarantee that its outputs respect
the reflection symmetry of the physical system. To impose this symmetry
exactly, we retain the same underlying network architecture but apply a
reflection-averaging projection to its outputs. Let
\[
    g_{\boldsymbol{\theta},1}
    (x,y,\widetilde{G},\widetilde{Wi}),
    \qquad
    g_{\boldsymbol{\theta},2}
    (x,y,\widetilde{G},\widetilde{Wi})
\]
denote the two unconstrained outputs of the network. The components of
the learned vector field are defined by
\begin{widetext}
\begin{equation}
\begin{aligned}
    f_{\boldsymbol{\theta},1}
    (x,y,\widetilde{G},\widetilde{Wi})
    &=
    \frac{1}{2}
    \Big[
        g_{\boldsymbol{\theta},1}
        (x,y,\widetilde{G},\widetilde{Wi})
        +
        g_{\boldsymbol{\theta},1}
        (x,-y,\widetilde{G},\widetilde{Wi})
    \Big],\\
    f_{\boldsymbol{\theta},2}
    (x,y,\widetilde{G},\widetilde{Wi})
    &=
    \frac{1}{2}
    \Big[
        g_{\boldsymbol{\theta},2}
        (x,y,\widetilde{G},\widetilde{Wi})
        -
        g_{\boldsymbol{\theta},2}
        (x,-y,\widetilde{G},\widetilde{Wi})
    \Big].
\end{aligned}
\label{eq:symmetry_projected_nn}
\end{equation}
\end{widetext}
By construction, the projected vector field satisfies the reflection symmetry conditions in
Eq.~\eqref{eq:reflection_parity_even_odd}
for any choice of the network parameters.
Thus, \(f_{\boldsymbol{\theta},1}\) is exactly even in \(y\), while
\(f_{\boldsymbol{\theta},2}\) is exactly odd in \(y\), independently of
the learned network parameters. This projection makes \(y=0\) an
invariant subspace and guarantees that every off-axis equilibrium
\((x^\ast,y^\ast)\) has a reflected equilibrium
\((x^\ast,-y^\ast)\) with the same local stability.

The symmetry-projected model is trained on the original equilibrium observations without reflected-data augmentation. Apart from the projection, the underlying network architecture, parameter scaling, and optimization settings are unchanged from the direct model. We use the same training objective as in
Eq.~\eqref{eq:direct_nn_total_loss}, with both loss terms evaluated using the projected vector field in Eq.~\eqref{eq:symmetry_projected_nn}. Stable equilibria are subsequently computed using the same multi-start root-solving procedure described in Appendix~\ref{app:neural_network_implementation}.

\begin{figure*}[t!]
    \centering
    \includegraphics[width=\textwidth]
    {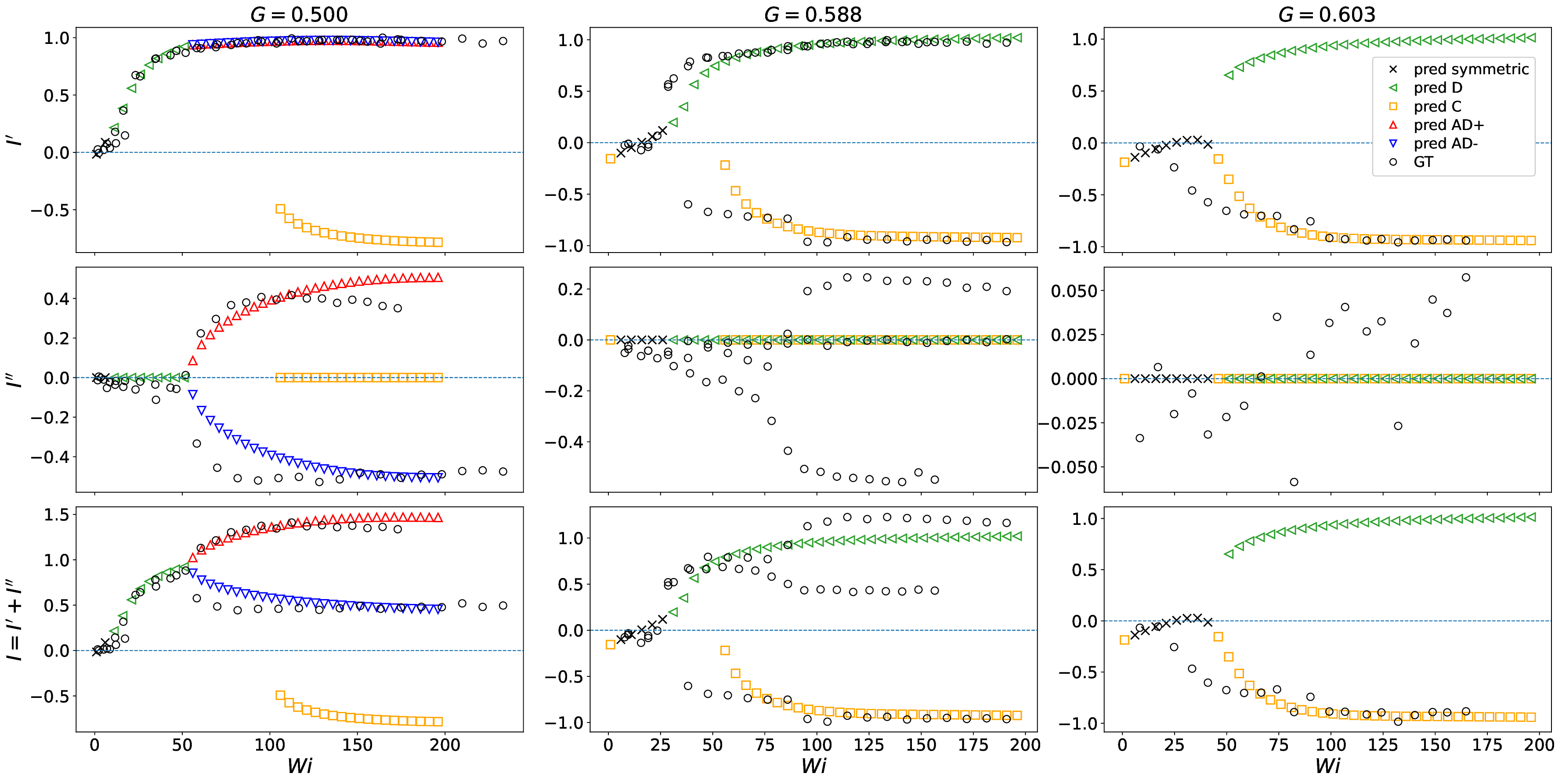}
   \caption{Bifurcation diagrams reconstructed from the
symmetry-projected neural-network normal form. The columns correspond
to \(G=0.500\), \(G=0.588\), and \(G=0.603\), while the rows show
\(I'\), \(I''\), and \(I=I'+I''\), respectively. Colored symbols
denote stable equilibria predicted by the learned vector field, and
black open circles denote the experimental measurements. The symmetry
projection produces paired AD\(^{+}\) and AD\(^{-}\) branches with
identical \(I'\) and opposite \(I''\). At \(G=0.500\), the model
captures the main D and AD branches but predicts an additional stable C
branch at large \(Wi\). At \(G=0.588\), the model approximately
reproduces the positive and negative \(I'\) branches, but the predicted
positive-\(I'\) branch remains on \(I''=0\) and does not split into the
two experimentally observed AD branches. At \(G=0.603\), the model
reproduces the high-\(Wi\) portion of the C branch but retains an
additional stable D branch.}
    \label{fig:symmetry_projected_nn_bifurcation}
\end{figure*}

Figure~\ref{fig:symmetry_projected_nn_bifurcation} shows the stable
equilibrium branches reconstructed from the symmetry-projected model.
At \(G=0.500\), the model reproduces the transition from the symmetric
state to D and the subsequent formation of two AD branches. In contrast
to the direct network and indeed by design, the two predicted AD branches have identical
\(I'\), opposite values of \(I''\), and the same stability. Their onset
occurs close to the experimentally observed transition. However, the
model also predicts an additional stable C branch for
\(Wi\gtrsim100\). At \(G=0.588\), the model approximately reproduces the positive and
negative \(I'\) branches but does not recover the corresponding
structure in \(I''\). The predicted positive-\(I'\) branch remains on
\(I''=0\) and therefore does not undergo the experimentally observed
transition from D to the two AD states. The model reproduces the
high-\(Wi\) portion of the C branch but not its continuation toward
lower \(Wi\). At \(G=0.603\), the high-\(Wi\) portion of the C branch is also
reproduced, but an additional stable D branch persists over the
moderate- and high-\(Wi\) ranges. Isolated C equilibria are additionally
predicted at low \(Wi\) for both \(G=0.588\) and \(G=0.603\). Thus, the
symmetry projection enforces the correct pairing of the AD states but
does not by itself eliminate unsupported stable branches or recover the
complete experimental branch structure.

\begin{figure*}[t!]
    \centering

    \begin{minipage}[t]{0.49\textwidth}
        \vspace{0pt}
        \centering
        \includegraphics[
            width=\linewidth,
            height=0.38\textheight,
            keepaspectratio
        ]{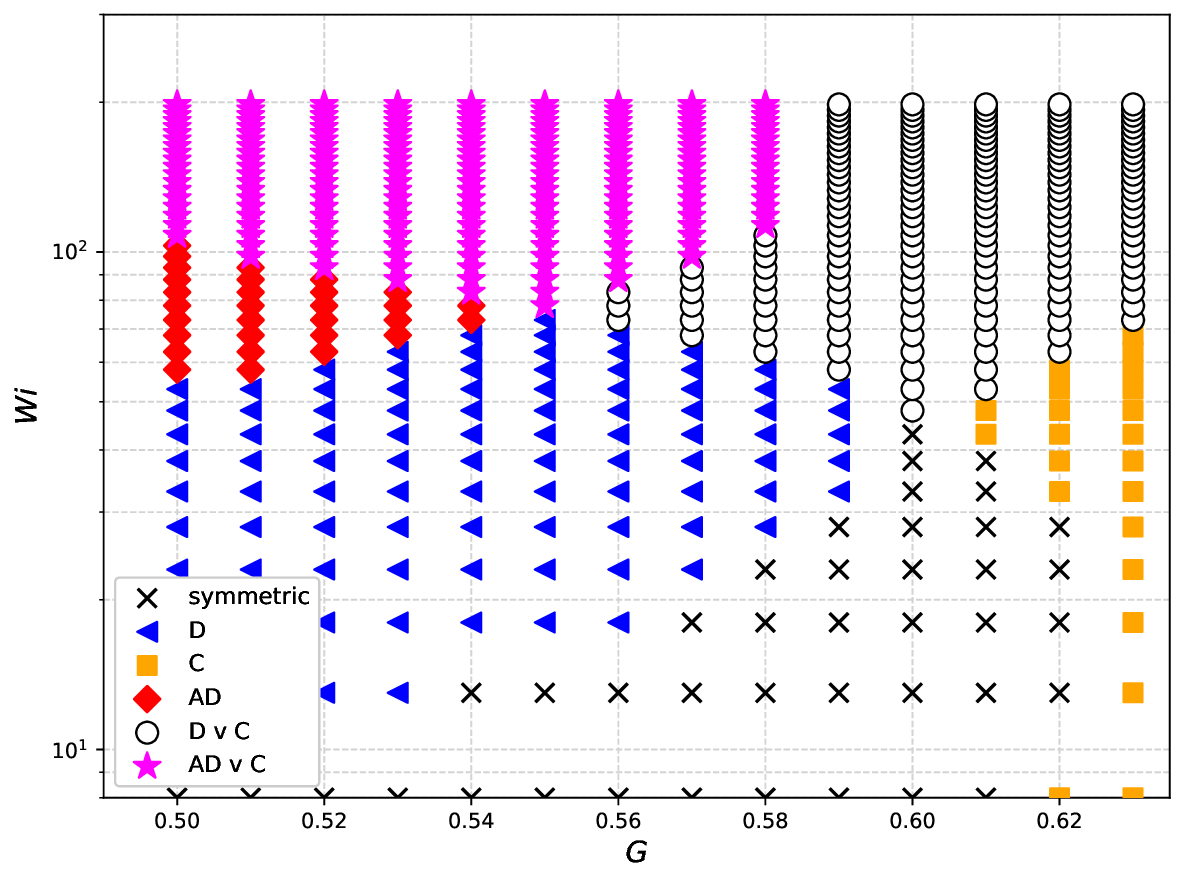}

        \vspace{0.3em}
        \textbf{(A)}
    \end{minipage}
    \hfill
    \begin{minipage}[t]{0.49\textwidth}
        \vspace{0pt}
        \centering
        \includegraphics[
            width=\linewidth,
            height=0.38\textheight,
            keepaspectratio
        ]{figures/flow_stability_diagram}

        \vspace{0.3em}
        \textbf{(B)}
    \end{minipage}

    \caption{Comparison of the reconstructed and experimental phase
    diagrams. (A) Phase diagram reconstructed from the stable
    equilibria of the symmetry-projected neural-network normal form.
    The symmetry projection ensures that the AD states occur as
    reflection-related pairs. However, an additional C branch produces
    an overly broad AD--C tristable region at small \(G\), the missing
    AD branches near \(G=0.588\) cause D--C bistability to persist at
    large \(Wi\), and an additional D branch extends the D--C region
    into the large-\(G\) regime. (B) Experimental flow-stability
    diagram in the \((G,Wi)\) parameter space. The symbols indicate the
    measured symmetric, D, C, and AD states, while the colored regions
    and dashed boundaries identify the corresponding stable and
    multistable regimes. The intermediate-gap region contains D--C
    bistability and AD--C tristability. {Reproduced from Hopkins et al.~\cite{PhysRevLett.126.054501}, with permission.}}
    \label{fig:symmetry_projected_nn_phase_diagram}
\end{figure*}

The corresponding phase diagram is shown in
Fig.~\ref{fig:symmetry_projected_nn_phase_diagram}(A). Compared with
the direct neural network, the phase classification is more regular
because the AD states now occur as reflection-related pairs. The model
captures the broad change from D and AD behavior at small \(G\) to
states involving C as \(G\) increases. However, comparison with the
experimental diagram in
Fig.~\ref{fig:symmetry_projected_nn_phase_diagram}(B) reveals several
systematic errors. For \(0.50\lesssim G\lesssim0.57\) and large \(Wi\), the additional C
branch coexists with the two AD branches. The model therefore predicts
AD--C tristability in a region where the experiments identify the AD
regime. Near \(G=0.588\), the predicted positive-\(I'\) branch remains
on \(I''=0\) instead of splitting into the two AD branches. As a result,
D--C bistability persists at large \(Wi\), replacing part of the
experimentally observed AD--C region. For \(G\gtrsim0.60\), the
additional D branch coexists with C and extends the predicted D--C
bistable region into the experimentally observed C regime. Thus, the
symmetry projection corrects the reflection-related pairing of the AD
states and improves the overall match with the experimentally observed
diagram, yet it does not fully determine the correct set of stable branches.

%The symmetry projection enforces the correct reflection symmetry but
%does not determine which symmetry-consistent equilibrium branches should
%exist. 
A key observation toward further improving our representation
of the normal form is the following:
the equilibrium and stability losses constrain the measured
states to be stable equilibria but do not penalize additional stable
roots away from the observations. To address this issue, we introduce
an iterative refinement procedure in which stable equilibria not
supported by the experimental data are identified and added as negative
samples during training.

\subsubsection{Iterative refinement with negative samples}
\label{sec:negative_samples}

%Although the symmetry projection enforces the required reflection
%structure, the equilibrium and stability losses constrain the vector
%field only at the measured states and do not exclude additional stable
%equilibria elsewhere. 
As shown in
Fig.~\ref{fig:symmetry_projected_nn_bifurcation}, the
symmetry-projected model predicts an additional stable C branch at
\(G=0.500\) and large \(Wi\), as well as an additional stable D branch
at \(G=0.603\) over the moderate- and high-\(Wi\) ranges. The former
produces AD--C tristability at small gap ratios, where the experiments
identify the AD regime, while the latter produces D--C bistability at
large gap ratios, where the experiments identify C. These errors are
shown in panel~(A) of
Fig.~\ref{fig:symmetry_projected_nn_phase_diagram}. 

To reduce these unsupported branches, we introduce an iterative refinement procedure that penalizes predicted stable branches not supported by the experimental observations. After each training stage, the stable branches are computed over the parameter
domain and compared with the experimental phase diagram. Predicted
stable roots whose flow-state classifications are inconsistent with the
experimental diagram are collected in
\begin{equation}
    \mathcal{D}_{-}
    =
    \left\{
        \left(
            x_j^{-},y_j^{-},G_j^{-},Wi_j^{-}
        \right)
    \right\}_{j=1}^{N_-}.
    \label{eq:negative_sample_dataset}
\end{equation}
Unlike the measured equilibria, these points are not constrained to
remain equilibria or locally stable during subsequent training.
Instead, the Jacobian at each negative sample is penalized unless it
has an eigenvalue with a sufficiently positive real part.

Let \(J_j^{-}\) denote the Jacobian evaluated at the \(j\)th negative
sample. We define the largest real part among its eigenvalues as
\begin{equation}
    \mu_{\max}(J_j^-)
    =
    \max_{\zeta\in\sigma(J_j^-)}
    \operatorname{Re}(\zeta),
    \label{eq:negative_max_real_eigenvalue}
\end{equation}
where \(\sigma(J_j^-)\) denotes the set of eigenvalues of \(J_j^-\).
A locally stable equilibrium satisfies \(\mu_{\max}<0\). We therefore
define the negative-sample loss as
\begin{equation}
    \mathcal{L}_{\mathrm{neg}}
    =
    \frac{1}{N_-}
    \sum_{j=1}^{N_-}
    \left[
        \max\left\{
            0,
            \gamma_{\mathrm{neg}}
            -
            \mu_{\max}(J_j^-)
        \right\}
    \right]^2,
    \label{eq:negative_stability_loss}
\end{equation}
where \(\gamma_{\mathrm{neg}}>0\) is an instability margin. This loss is
active whenever
\(\mu_{\max}(J_j^-)<\gamma_{\mathrm{neg}}\) and encourages at least one
Jacobian eigenvalue to have a real part larger than
\(\gamma_{\mathrm{neg}}\). Because no equilibrium residual is imposed
at the negative samples, a corresponding root may move or disappear
during fine-tuning. If it remains an equilibrium, the loss encourages
it to become locally unstable.

The model is fine-tuned using
\begin{equation}
    \mathcal{L}_{\mathrm{iter}}
    =
    \mathcal{L}_{\mathrm{eq}}
    +
    \lambda_{\mathrm{stab}}
    \mathcal{L}_{\mathrm{stab}}
    +
    \lambda_{\mathrm{neg}}
    \mathcal{L}_{\mathrm{neg}}.
    \label{eq:iterative_training_loss}
\end{equation}
The first two terms maintain the equilibrium and stability constraints
at the measured states, while the third discourages the selected
negative samples from remaining stable equilibria. After each
fine-tuning stage, the equilibrium branches are recomputed, and
additional unsupported stable roots can be included in the
negative-sample pool in the next refinement step. The
symmetry-projected architecture is retained throughout, and no
reflected-data augmentation is used. The negative-sample region,
classification criteria, loss weights, and fine-tuning settings are
reported in Appendix~\ref{app:negative_sample_implementation}.

\begin{figure*}[t!]
    \centering
    \includegraphics[width=\textwidth]
    {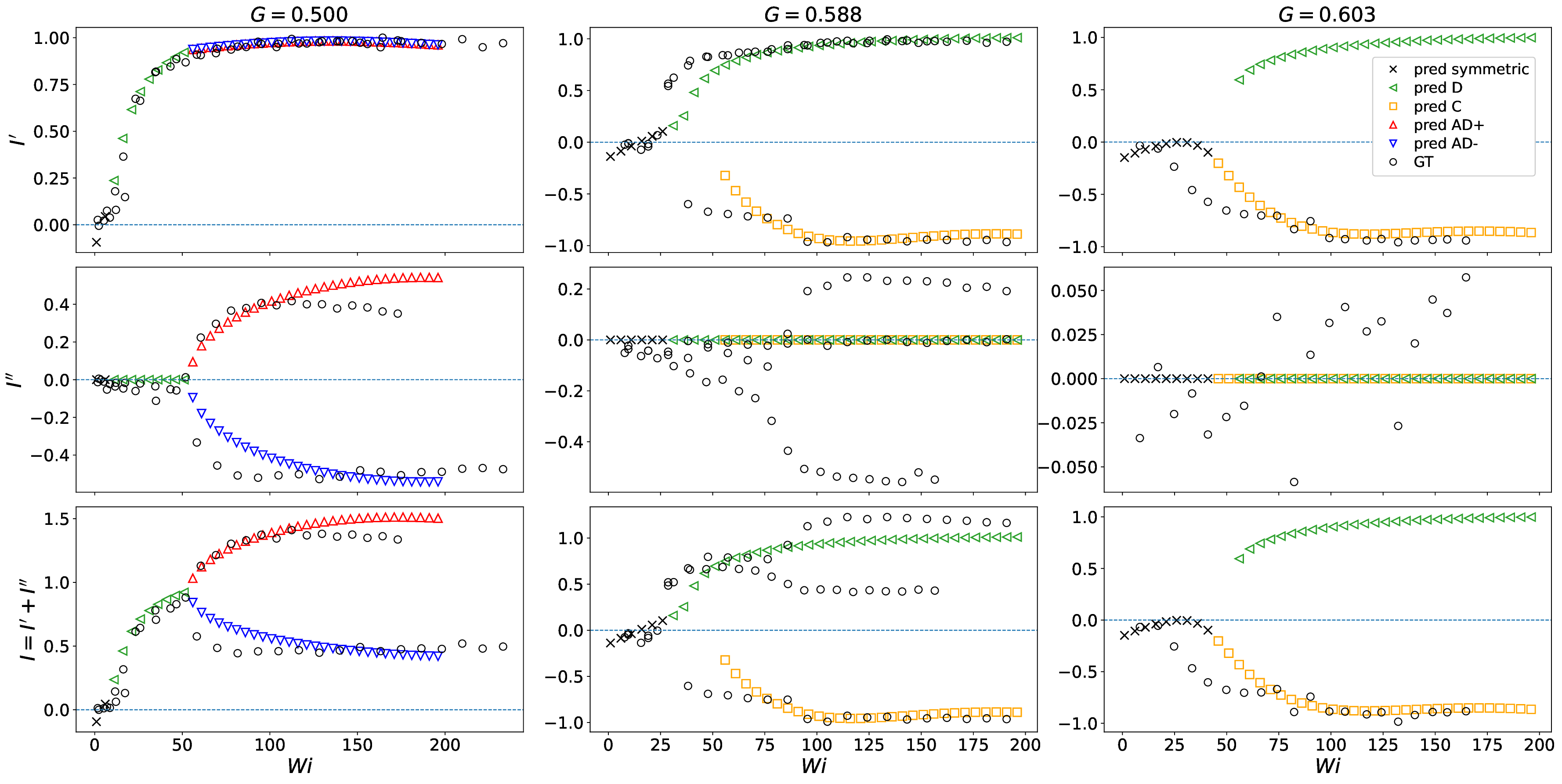}
    \caption{Bifurcation diagrams obtained after iterative refinement
    using negative samples from the additional C branch at \(G=0.500\).
    The columns correspond to \(G=0.500\), \(G=0.588\), and
    \(G=0.603\), while the rows show \(I'\), \(I''\), and
    \(I=I'+I''\), respectively. Colored symbols denote predicted stable
    equilibria, and black open circles denote the experimental
    measurements. The targeted C branch at \(G=0.500\) is removed,
    while the D and reflection-related AD branches at \(G=0.500\) and
    the C branches at larger \(G\) are preserved. The result
    demonstrates that negative-sample refinement can selectively modify
    the learned equilibrium structure without breaking the imposed
    reflection symmetry.}
    \label{fig:negative_c_samples_bifurcation}
\end{figure*}

We apply this procedure to the additional C branch predicted at
\(G=0.500\). As shown in
Fig.~\ref{fig:negative_c_samples_bifurcation}, the targeted branch is
removed over the sampled \(Wi\) range, while the D branch and the two
reflection-related AD branches are preserved. The C branches at
\(G=0.588\) and \(G=0.603\) also remain, showing that the refinement
does not eliminate C behavior throughout the parameter domain. The
other predicted branches change little after fine-tuning. These results
show that negative samples can selectively remove an unsupported stable
branch while preserving the reflection symmetry and the experimentally
supported branches elsewhere.

\begin{figure*}[t!]
    \centering

    \begin{minipage}[t]{0.49\textwidth}
        \vspace{0pt}
        \centering
        \includegraphics[
            width=\linewidth,
            height=0.38\textheight,
            keepaspectratio
        ]{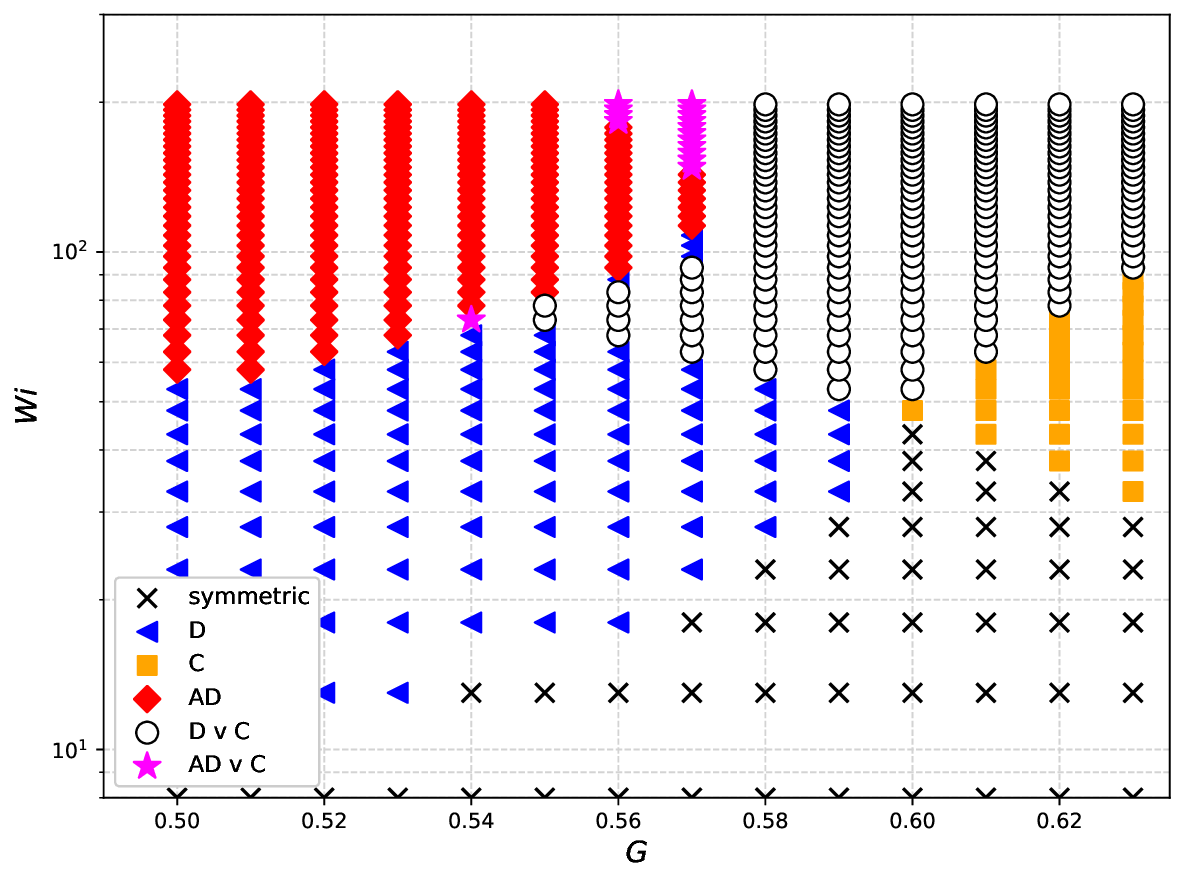}

        \vspace{0.3em}
        \textbf{(A)}
    \end{minipage}
    \hfill
    \begin{minipage}[t]{0.49\textwidth}
        \vspace{0pt}
        \centering
        \includegraphics[
            width=\linewidth,
            height=0.38\textheight,
            keepaspectratio
        ]{figures/flow_stability_diagram}

        \vspace{0.3em}
        \textbf{(B)}
    \end{minipage}

    \caption{Comparison of the reconstructed and experimental phase
    diagrams. (A) Phase diagram reconstructed after iterative
    refinement using negative samples from the additional C branch.
    Removing this branch recovers the high-\(Wi\) AD region at small
    \(G\) and confines most of the predicted AD--C tristability to
    intermediate gap ratios. At intermediate and large \(G\), the
    missing AD states near \(G=0.588\) and the additional D branch at
    larger \(G\) produce a broader D--C bistable region than observed
    experimentally. (B) Experimental flow-stability diagram in the
    \((G,Wi)\) parameter space. The symbols indicate the measured
    symmetric, D, C, and AD states, while the colored regions and dashed
    boundaries identify the corresponding stable and multistable
    regimes. The intermediate-gap region contains D--C bistability and
    AD--C tristability. Reproduced from Hopkins et al.~\cite{PhysRevLett.126.054501}, with permission.}
    \label{fig:negative_c_samples_phase_diagram}
\end{figure*}

The corresponding phase diagram is shown in
Fig.~\ref{fig:negative_c_samples_phase_diagram}(A). The branch-level
correction in Fig.~\ref{fig:negative_c_samples_bifurcation} produces a
clear improvement in the phase classification. At small \(G\), the
model now recovers the progression from the symmetric state to D and
then AD as \(Wi\) increases. Removing the additional C branch eliminates
most of the previously overpredicted AD--C region and restores the AD
regime over the small-\(G\), high-\(Wi\) domain. The predicted AD--C
region is consequently concentrated near the intermediate gap ratios,
in closer agreement with the experimental diagram in
Fig.~\ref{fig:negative_c_samples_phase_diagram}(B). The C states at
larger \(G\) are also preserved.

The remaining differences are concentrated at intermediate and large
\(G\). Near \(G=0.588\), the predicted positive-\(I'\) branch remains
on \(I''=0\), causing D--C bistability to persist at large \(Wi\)
instead of producing the observed AD--C tristability. For
\(G\gtrsim0.60\), the additional D branch extends the predicted D--C
region into part of the experimentally observed C regime. Nevertheless,
the refined phase diagram captures the small-\(G\) AD region much more
accurately while retaining the C-containing regimes at larger \(G\).

Note that we also applied the same refinement procedure to the additional D
branch at \(G=0.603\). However, the branch could not be removed without
degrading the fit to the measured equilibria. The predicted D branch
may be an artifact of the learned model or a stable state that was not
observed experimentally. The available data are insufficient to
distinguish between these possibilities. Additional experimental measurements would be needed to determine whether this branch exists in the physical system.

The successful refinement of the C branch at \(G=0.500\) demonstrates
that negative samples can incorporate information about the absence of
a stable branch while preserving the measured equilibria and the
imposed reflection symmetry. Together, the symmetry projection and
negative-sample refinement provide direct control over both the
symmetry and the stable equilibrium structure of the learned vector
field.

\begingroup
\color{black}

\subsection{Quantitative comparison}
\label{sec:quantitative_comparison}

To compare the polynomial and neural-network normal forms in a quantitative
fashion, we measure the distance between each experimentally observed equilibrium and the nearest stable equilibrium predicted by the model at the same control parameters. The equilibrium reconstruction error is defined as
\begin{equation}
    E_{\mathrm{eq}} =
    \frac{1}{N_{\mathrm{eq}}}
    \sum_{i=1}^{N_{\mathrm{eq}}}
    \min_{\boldsymbol{\xi}^{*}\in\mathcal{S}(G^i,Wi^i)}
    \left\|
    \boldsymbol{\xi}^i-\boldsymbol{\xi}^{*}
    \right\|_2,
    \label{eq:equilibrium_error}
\end{equation}
where $\boldsymbol{\xi}^i=(x^i,y^i)$ denotes the experimentally observed equilibrium and
$\mathcal{S}(G^i,Wi^i)$ is the set of stable equilibria predicted by the model at the
corresponding parameters. The error is evaluated using all $N_{\mathrm{eq}}=N=445$ experimental equilibrium observations, with multiple equilibria possibly corresponding to the same $(G,Wi)$.

\begin{table}[t]
\caption{
Equilibrium reconstruction errors for the polynomial
and neural-network normal forms, evaluated using
445 experimental equilibrium observations.
}
\label{tab:quant_comparison}
\centering
\footnotesize
\setlength{\tabcolsep}{4pt}
\renewcommand{\arraystretch}{1.15}
\begin{tabular}{@{}lc@{}}
\hline\hline
Model & $E_{\mathrm{eq}}$ \\
\hline
Polynomial (unrestricted) & 0.3544 \\
Polynomial (even--odd) & 0.2960 \\
NN (direct) & 0.1814 \\
NN (symmetry-projected) & \textbf{0.1503} \\
\shortstack[l]{NN (symmetry-projected\\+ negative samples)}
    & 0.2067 \\
\hline\hline
\end{tabular}
\end{table}

The results are summarized in Table~\ref{tab:quant_comparison}. Overall, the NN models give smaller equilibrium reconstruction errors than the polynomial models. Imposing reflection symmetry reduces $E_{\mathrm{eq}}$ from $0.3544$ to $0.2960$ for the polynomial models and from $0.1814$ to $0.1503$ for the NN models. After negative-sample refinement, the error for the symmetry-projected NN is $0.2067$. 
While this seems like a (slight) increase of the reconstruction error
in comparison to non-NN methods, the important advantage of the negative samples
strategy is its improved qualitative correspondence with the experimental 
findings, including as concerns the characterization of multistable regions.
Since $E_{\mathrm{eq}}$ measures only the distance to the nearest stable equilibrium, it does not describe the full set of predicted stable branches, which are examined in the bifurcation and phase
diagrams above.

\endgroup

\section{Conclusions and Outlook}
\label{sec:conclusions}

In this work, we investigated whether sparse measurements of stable
flow states can be used to learn a low-dimensional, reduced dynamical
system for viscoelastic flow past two side-by-side microcylinders. The
learned vector field couples the two flow asymmetry parameters \(I'\)
and \(I''\), with the gap ratio \(G\) and the Weissenberg number \(Wi\)
serving as control parameters. Its stable equilibria provide a unified
description of the measured branches and phase diagram over the
\((G,Wi)\) parameter domain, rather than separate local fits for the
individual transitions.

A central result is that equilibrium locations and local stability
alone do not suffice ---at least in the case at hand--- in order to determine the global bifurcation structure. The polynomial
model provided an interpretable representation but produced shifted
transitions and additional stable branches. A more flexible direct
neural network reproduced several measured branches but did not
preserve the reflection relation between the two AD states. Imposing
the reflection symmetry directly through the network architecture
resolved this inconsistency: every predicted AD state has a counterpart
with the same \(I'\), opposite \(I''\), and the same local stability.
The symmetry constraint determines the relation between these states,
but it does not determine which stable branches should exist elsewhere
in the domain. The quantitative comparison also shows that the neural-network
models yield smaller equilibrium reconstruction errors than
the polynomial models.

We addressed this remaining ambiguity through iterative refinement with
negative samples. Stable roots that were inconsistent with the
experimental phase diagram were identified after training and
discouraged from remaining stable equilibria during fine-tuning. This
procedure removed the additional C branch at small \(G\) while
preserving the D and AD branches at \(G=0.500\) and the C branches at
larger \(G\). The resulting phase diagram recovered the symmetric--D--AD
sequence over most of the small-\(G\) region and substantially reduced
the previously overpredicted AD--C tristable regime. These results show that the reflection symmetry and the negative samples provide different
but complementary pieces of information: the former fixes the relation between
symmetry-related states, while the latter helps select the stable
branches supported by the experimental observations.

The final reconstruction still differs from the experimental diagram in
two regions. Near \(G=0.588\), the positive-\(I'\) branch remains on
\(I''=0\), and the transition from D to the two AD states is not
recovered. Near \(G=0.603\), an additional stable D branch remains.
Attempts to remove this branch degraded the fit at the measured
equilibria. The available data cannot determine whether the branch is
an artifact of the learned model or a stable state that was not
observed experimentally. {This uncertainty highlights the limitations of learning a global vector field from stable equilibrium measurements alone. Further experiments could help determine whether the additional predicted branch exists but was not observed, possibly because of a small basin of attraction or different initial conditions.}

{Transient measurements and responses to controlled perturbations could provide additional information about the dynamics beyond the observed equilibria. A more systematic approach to selecting negative samples could also improve the reconstruction of the phase diagram. Additional measurements at intermediate and large $G$ would help clarify the missing AD
transition and the predicted D branch. Finally, the proposed framework could be extended to more complex geometries, including larger obstacle arrays
and regular lattices. Such extensions are currently being considered for future work.}

\begin{acknowledgments}
The authors are especially grateful to Professor Wei Zhu (Georgia Tech)
for his guidance and substantial feedback throughout this
research thread.
{SJH and AQS gratefully acknowledge the support of the Okinawa Institute of Science and Technology Graduate University (OIST) with subsidy funding from the Cabinet Office, Government of Japan. SJH and AQS also acknowledge Kakenhi funding from the Japan Society for the Promotion of Science (JSPS, Grant Nos. 24K07332,and 24K00810).}
%P.G.K was supported in part by the U.S. National Science Foundation under the award PHY-2408988. 
This material is partially based upon work supported by the U.S. National Science Foundation under the award PHY-2408988 (PGK).
This research was partly conducted while P.G.K. was  visiting the Okinawa Institute of Science and
Technology (OIST) through the Theoretical Sciences Visiting Program (TSVP), the University of
Sydney through the visitor program of the Sydney Mathematical Research Institute (SMRI) and the Department of Mechanical Engineering at Seoul National
University through a Fulbright Fellowship. Their support is gratefully acknowledged.
Finally, this work was also  supported by a grant from the Simons Foundation [SFI-MPS-SFM-00011048, P.G.K].
\end{acknowledgments}

\section*{Code Availability}

The code used for model training, equilibrium reconstruction, phase
classification, and figure generation is available in the
\href{https://github.com/Shaoxuan-Chen/Learning-Viscoelastic-Flow-Normal-Form}
{GitHub repository}.

% \newpage
% \appendix
% \onecolumn
\appendix

% ```latex

\section{Polynomial implementation details}
\label{app:polynomial_implementation}

The final polynomial model is trained using all original
equilibrium observations without reflected-data augmentation.
Since the even--odd symmetry is imposed exactly, each reflected
observation contributes the same equilibrium and stability
losses as its original counterpart. Therefore, including the
reflected observations would leave the full-batch averaged
training objective unchanged.

The input scaling is defined in
Eq.~\eqref{eq:polynomial_scaling}.
For the experimental data considered here, none of the
scaling factors is close to zero, avoiding potential numerical
instabilities associated with division by small values.
All polynomial features and stability constraints are evaluated
in the scaled coordinates.

The polynomial dictionaries follow the degree constraints
in Eq.~\eqref{eq:polynomial_degree_constraints} and the
reflection symmetry in Eq.~\eqref{eq:reflection_parity_even_odd}.
With the constant term excluded, the resulting dictionaries
contain \(44\) candidate features for \(f_1\) and \(30\)
for \(f_2\).

The coefficient vectors \(\mathbf{a}\) and \(\mathbf{b}\)
are initialized at zero and optimized in \(64\)-bit
floating-point precision using full-batch Adam.
The training objective is given in
Eq.~\eqref{eq:polynomial_total_loss}, with the stability
penalty defined in Eq.~\eqref{eq:stability_loss}.
The hyperparameters are
\begin{equation}
\begin{aligned}
    &\lambda_{\mathrm{stab}}=500,
    \quad
    \lambda_{\mathrm{sparse}}=10^{-5},
    \quad
    \alpha=0.03,\\
    &\beta=10^{-5},
    \quad
    w_{\det}=100.
\end{aligned}
\end{equation}

Training is performed for \(25\,000\) optimization steps.
The initial learning rate is \(5\times10^{-4}\) and is
multiplied by \(0.2\) at steps \(12\,000\) and \(20\,000\).
Random seeds are fixed for reproducibility.

After optimization, the polynomial coefficients are
converted for equilibrium solving using
Eq.~\eqref{eq:polynomial_coefficient_conversion}.
To express the corresponding vector field in the original
state coordinates, let
\(\mathbf{z}=(x,y)^{\mathsf T}\),
\(\widetilde{\mathbf{z}}=S^{-1}\mathbf{z}\), and
\(S=\operatorname{diag}(s_x,s_y)\).
The coordinate transformation gives
\begin{equation}
    \mathbf{F}_{\mathrm{raw}}(\mathbf{z},G,Wi)
    =
    S\mathbf{F}_{\mathrm{scaled}}
    (\widetilde{\mathbf{z}},G/s_G,Wi/s_{Wi}),
\end{equation}
where \(\mathbf{F}_{\mathrm{scaled}}=(f_1,f_2)^{\mathsf T}\)
is the learned vector field in the scaled coordinates.
The coefficients of the two components therefore satisfy
\begin{equation}
    c_{\mathrm{raw},1}=s_xc_{\mathrm{eq},1},
    \qquad
    c_{\mathrm{raw},2}=s_yc_{\mathrm{eq},2}.
\end{equation}

The Jacobians in the scaled and original coordinates satisfy
\begin{equation}
    J_{\mathrm{raw}}
    =
    S J_{\mathrm{scaled}}S^{-1},
    \qquad
    J_{\mathrm{scaled}}
    =
    \frac{\partial\mathbf{F}_{\mathrm{scaled}}}
    {\partial\widetilde{\mathbf{z}}},
\end{equation}
and therefore have identical eigenvalues and local stability.
Stability is evaluated using \(J_{\mathrm{scaled}}\) as in training. 

For Bertini, terms with coefficient magnitude below \(10^{-10}\) are omitted.
The residual of each computed root is evaluated
using the full polynomial equations.

% ```latex
\section{Polynomial ablation study}
\label{app:polynomial_ablation}

To assess the role of the imposed reflection symmetry, we compare the
final symmetry-restricted polynomial model with a fully unrestricted
polynomial representation. The unrestricted model uses the same degree
bounds, loss functions, optimization settings, and equilibrium-solving
procedure described in
Appendix~\ref{app:polynomial_implementation}, but no restriction is
placed on the parity of \(y\) in either component of the vector field.
The model is fitted using the original equilibrium observations without
reflected-data augmentation.

\begin{figure*}[htbp]
    \centering
    \includegraphics[width=.95\textwidth]
    {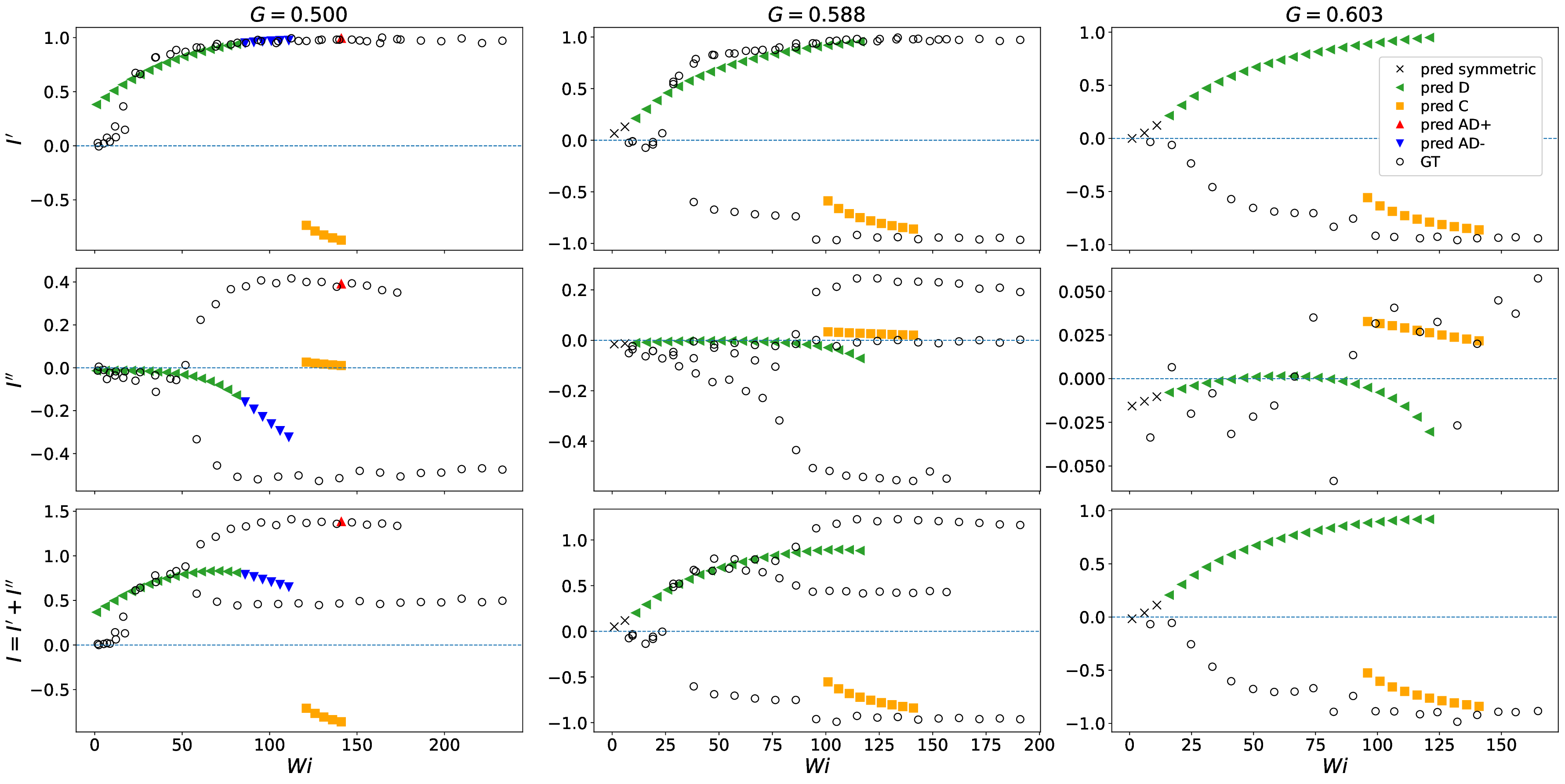}
    \caption{Bifurcation diagrams obtained from the unrestricted
    polynomial model. The columns correspond, from left to right, to
    \(G=0.500\), \(G=0.588\), and \(G=0.603\), while the rows show
    \(I'=x\), \(I''=y\), and \(I=I'+I''\), respectively. Black open
    circles denote the experimental equilibrium measurements, and
    colored symbols denote the stable equilibria predicted by the
    learned vector field. Predicted equilibria satisfying
    \(\lvert I'\rvert<0.15\) and \(\lvert I''\rvert<0.15\) are classified
    as symmetric. Without the even--odd restriction, the model does not
    preserve the invariant subspace \(I''=0\) or generate the two AD
    states as a reflection-related pair. No reflected-data augmentation
    is used.}
    \label{fig:polynomial_unrestricted_branches}
\end{figure*}

Figure~\ref{fig:polynomial_unrestricted_branches} compares the predicted
and experimental equilibrium branches at \(G=0.500\), \(G=0.588\), and
\(G=0.603\). Although the unrestricted model captures parts of the
observed variation in \(I'\), it does not reproduce the required
reflection structure in \(I''\). At \(G=0.500\), the model generates
only one of the two AD branches and additionally predicts a C branch at
large \(Wi\). The reflection-related AD pair is also not recovered at
\(G=0.588\). At \(G=0.603\), the predicted C and D branches exhibit
systematic nonzero values of \(I''\), rather than remaining on the
invariant subspace \(I''=0\). These results show that, for the
unrestricted polynomial representation, the equilibrium and stability
constraints at the measured states are insufficient to recover the
physical reflection symmetry of the vector field.

Because the unrestricted model already exhibits substantial
discrepancies at the level of its equilibrium branches, a separate
phase diagram is not reported for this ablation. The complete
bifurcation and phase-diagram reconstruction obtained from the final
symmetry-restricted polynomial model is presented in
Sec.~\ref{sec:polynomial_normal_form}.

\section{Direct neural-network implementation details}
\label{app:neural_network_implementation}

The direct neural-network model is fitted using all available original
equilibrium observations without reflected-data augmentation or a
train--test split. Only the control parameters are scaled, following
Eq.~\eqref{eq:polynomial_scaling}, while the state variables \(x\)
and \(y\) are left unscaled. The network
architecture is described in Sec.~\ref{sec:direct_neural_network}. All
computations are performed in \(64\)-bit floating-point precision, and fixed random seeds are used to ensure
reproducibility.

The model is optimized in full batches using AdamW with zero weight
decay. The training objective is given in
Eq.~\eqref{eq:direct_nn_total_loss}, with the stability penalty
defined in Eq.~\eqref{eq:stability_loss}. The hyperparameters are
\begin{equation}
    \lambda_{\mathrm{stab}}=50,
    \quad
    \alpha=0.09,
    \quad
    \beta=10^{-3},
    \quad
    w_{\det}=50.
\end{equation}

Training is performed for \(60\,000\) optimization steps. The initial
learning rate is \(5\times10^{-4}\) and is reduced by a factor of
\(0.5\) after \(20\,000\) and \(40\,000\) steps. Consequently, the
learning rates in the three training intervals are
\(5\times10^{-4}\), \(2.5\times10^{-4}\), and
\(1.25\times10^{-4}\), respectively.

For each fixed parameter pair \((G,Wi)\), the equilibrium equations are solved using the Powell hybrid algorithm implemented in
\texttt{scipy.optimize.root}. The solver is initialized from a uniform
\(9\times9\) grid over
\[
    (x_0,y_0)\in[-1.1,1.1]\times[-1.1,1.1],
\]
giving \(81\) initial guesses at each parameter pair. A converged solution \((x^\ast,y^\ast)\) is accepted when
\begin{equation}
    \sqrt{
        \left|
        f_{\boldsymbol{\theta},1}
        (x^\ast,y^\ast,\widetilde{G},\widetilde{Wi})
        \right|^2
        +
        \left|
        f_{\boldsymbol{\theta},2}
        (x^\ast,y^\ast,\widetilde{G},\widetilde{Wi})
        \right|^2
    }
    <10^{-6}.
    \label{eq:appendix_direct_nn_residual}
\end{equation}

Because different initial guesses may converge to the same equilibrium,
two solutions are treated as identical when their Euclidean distance in
the \((x,y)\) plane is less than \(10^{-4}\). The Jacobian is evaluated
by automatic differentiation, and the stability of each equilibrium
is determined using Eq.~\eqref{eq:stability_conditions}.

Stable equilibria are classified into the different flow states
using the criteria defined in Sec.~\ref{sec:general_setting},
with \(\delta=0.15\). 
For the symmetry-constrained polynomial and neural-network
models, reflection-related stable roots are identified
at each fixed \((G,Wi)\). Two stable roots initially classified as D are identified
as a reflection-related AD pair if their \(x\) coordinates are nearly identical and their \(y\) coordinates are approximately opposite and sufficiently separated. Specifically, we require
\[
    |x_1-x_2|<10^{-3},
    \qquad
    |y_1+y_2|<10^{-3},
\]
together with
\[
    y_1y_2<0,
    \qquad
    |y_1-y_2|>10^{-3}.
\]
The roots are then labeled AD\(^{+}\) and AD\(^{-}\) according to the signs of their \(y\) coordinates. Stable roots that do not satisfy any of these criteria are labeled as \emph{other}. 

The bifurcation diagrams are constructed at
\[
    G=0.500,\qquad G=0.588,\qquad G=0.603
\]
by solving the equilibrium equations at
\(Wi=1,6,\ldots,196\). The two-parameter phase diagram is evaluated on
the grid
\begin{equation}
    G=0.50,0.51,\ldots,0.63,
    \qquad
    Wi=8,13,\ldots,198.
\end{equation}
For the phase classification, only stable roots satisfying
\(|x|\leq1.1\) and \(|y|\leq1.1\) are retained. Parameter values containing both D and C equilibria are classified as D--C bistable,
while those containing a C equilibrium together with AD equilibria are
classified as AD--C multistable.

\section{Negative-sample refinement details}
\label{app:negative_sample_implementation}

For the negative-sample refinement, we follow the symmetry-projected neural-network framework described in Sec.~\ref{sec:symmetry_neural_network}. In these experiments, we first train a network with three fully connected hidden layers of width \(50\), with hyperbolic tangent activations, followed by two raw outputs. Reflection averaging is then applied as in Eq.~\eqref{eq:symmetry_projected_nn}
to enforce the required even--odd structure. At each refinement iteration, stable equilibria are recomputed over
\begin{equation}
    G=0.50,0.51,\ldots,0.55,
    \qquad
    Wi=100,105,\ldots,200.
    \label{eq:appendix_negative_search_region}
\end{equation}
The same initialization and convergence criteria as in Appendix~\ref{app:neural_network_implementation}
are used, with only stable roots within the search
domain retained. Among these roots, those classified as C and satisfying
\[
    -1.2<x<0,
    \qquad
    |y|<0.2
\] are retained as negative samples. Newly detected roots are added to a
cumulative negative-sample pool, and exact duplicates in
\((G,Wi,x,y)\) are removed. If no negative C roots are detected, the
refinement terminates early.

For each negative sample, the largest real part of the Jacobian eigenvalues is computed from its trace and determinant. Defining
\begin{equation}
    \Delta_j
    =
    \operatorname{tr}(J_j^-)^2
    -
    4\det(J_j^-),
\end{equation}
the largest eigenvalue real part is computed as

\begin{equation}
    \mu_{\max}(J_j^-)
    =
    \begin{cases}
        \displaystyle
        \frac{
            \operatorname{tr}(J_j^-)
            +
            \sqrt{\Delta_j}
        }{2},
        &\Delta_j\geq0,\\[3mm]
        \displaystyle
        \frac{\operatorname{tr}(J_j^-)}{2},
        &\Delta_j<0.
    \end{cases}
    \label{eq:appendix_negative_max_eigenvalue}
\end{equation}

In the numerical implementation, the argument of the square root is
clipped below at zero and a regularization of \(10^{-12}\) is added.

The negative-sample loss is defined in
Eq.~\eqref{eq:negative_stability_loss},
with \(\gamma_{\mathrm{neg}}=10^{-4}\). Only the Jacobian is used in this term; no equilibrium loss is imposed
on the negative samples. Consequently, fine-tuning may move or remove
the corresponding root, or retain it as an unstable equilibrium. 

The network is fine-tuned using the objective defined in
Eq.~\eqref{eq:iterative_training_loss}.
The equilibrium and stability loss parameters remain unchanged from Appendix~\ref{app:neural_network_implementation},
and we set \(\lambda_{\mathrm{neg}}=1\). Three negative-sample iterations are performed. During each iteration,
the current network parameters are fine-tuned for \(2000\) full-batch
optimization steps using AdamW with learning rate \(10^{-4}\) and zero
weight decay. The network is not reinitialized between iterations.
After each fine-tuning stage, the stable equilibria are recomputed, and
any remaining C roots in the region
Eq.~\eqref{eq:appendix_negative_search_region} are added to the
negative-sample pool. 

\bibliography{main}

\end{document}